\documentclass[aps,pra,twocolumn,groupedaddress,floatfix,superscriptaddress]{revtex4-2}

\usepackage[english]{babel}
\usepackage{bbold}
\usepackage{enumitem}
\usepackage[latin9]{inputenc}
\usepackage{braket}
\usepackage{amsmath}
\usepackage{amssymb}
\usepackage{amsfonts}
\usepackage{color}
\usepackage{graphicx}
\usepackage{comment}

\usepackage[colorlinks=true,
            linkcolor=magenta,
            urlcolor=blue,
            citecolor=blue]{hyperref}

\newcommand{\mb}[1]{\mathbf{ #1}}
\newcommand{\bs}[1]{\boldsymbol{ #1}}
\newcommand{\mr}[1]{\mathrm{ #1}}

\newcommand{\hatboldL}{\hspace{-0.085cm}\hat{\hspace{0.085cm} \mathbf{L}}}

\emergencystretch=\maxdimen
\begin{document}
\title{High-dimensional SO(4)-symmetric Rydberg manifolds for quantum simulation}
%\title{Rydberg atoms in the SO(4)-symmetric manifold: quantum simulation Hilbert space of quantum simulators with}
%\title{A Rydberg quantum simulator with large angular momenta}
\author{Andreas Kruckenhauser}
\email{andreas.kruckenhauser@uibk.ac.at}
\author{Rick van Bijnen}
\author{Torsten V. Zache}
\author{Marco Di Liberto}
\author{Peter Zoller}
\affiliation{Institute for Quantum Optics and Quantum Information of the Austrian Academy of Sciences, Innsbruck, Austria}
\affiliation{Institute for Theoretical Physics, University of Innsbruck, Innsbruck, Austria}

\begin{abstract}

We develop a toolbox for manipulating arrays of Rydberg atoms prepared in high-dimensional hydrogen-like manifolds in the regime of linear Stark and Zeeman effect.
We exploit the SO(4) symmetry to characterize the action of static electric and magnetic fields as well as microwave and optical fields on the well-structured manifolds of states with principal quantum number $n$. 
This enables us to construct generalized \emph{large-spin} Heisenberg models for which we develop state-preparation and readout schemes.
Due to the available large internal Hilbert space, these models provide a natural framework for the quantum simulation of Quantum Field Theories, which we illustrate for the case of the sine-Gordon and massive Schwinger models.
Moreover, these high-dimensional manifolds also offer the opportunity to perform quantum information processing operations for qudit-based quantum computing, which we exemplify with an entangling gate and a state-transfer protocol for the states in the neighborhood of the circular Rydberg level.

\end{abstract}

\maketitle

%%%%%%%%%%%%%%%   Introduction   %%%%%%%%%%%%%%%%

\section{Introduction}
\label{sec:intro}

In the early days of quantum mechanics, Wolfgang Pauli proposed an elegant algebraic solution of the Hydrogen problem based on the SO(4) symmetry of the Hamiltonian for an electron moving in a Coulomb potential \cite{Pauli1926}. This provided not only a derivation of quantized energy levels for Hydrogen described by the Rydberg formula $E_{n}= -{\rm Ry}/n^2 $, with $n=1,2,\ldots$ the principal quantum number, but also explained the $n^2$-fold (orbital) degeneracy of the $n$-manifolds. 
In (weak) static external electric and magnetic fields, these degeneracies are lifted resulting in a linear Stark and Zeeman effect~\cite{Hulet1983, Delande1988NewMethod}, leading to a large and regularly structured internal state space associated with each value of $n$. 

Here we look at such high-dimensional and well-structured internal atomic state spaces, as provided by hydrogen-like Rydberg states of multi-electron atoms, as a novel opportunity for both storing and manipulating quantum information for quantum- simulation and computing.
This involves the control of a given $n$ manifold of single Rydberg atoms with external electromagnetic fields, as well as entangling operations via strong dipolar interactions. 
Availability and control of large internal Hilbert spaces is of interest in quantum simulation of general scalar field theories \cite{jordan2012quantum}, gauge theories of high-energy physics \cite{banuls2020simulating}, condensed matter models with large spins \cite{sachdev2011quantum, auerbach2012interacting, Gorshkov2010, Taie2012, Patscheider2020, Chomaz2022}, or the emulation of synthetic spatial dimensions \cite{Celi2014, Mancini2015, Chalopin2020, Ozawa2019, Lepoutre2019}. 
In view of the recent experimental advances with neutral atoms laser-excited to Rydberg states \cite{Browaeys2020,Morgado2021}, we will explore some of these opportunities from a theory perspective and with Pauli's algebraic formalism providing a framework to treat the single- and many-body problem.

Rydberg states are already known as one of the most promising neutral-atom platforms for quantum- information and simulation purposes in various scenarios \cite{Altman2021,Browaeys2020,Morgado2021}.
These include the storage of qubits (or qudits) in ground states where fast and high fidelity entangling gates can be performed via interactions between atoms that are laser-coupled to Rydberg states \cite{Jaksch2000, Lukin2001, Wilk2010, Isenhower2010, Jau2016, Levine2019, Graham2019, Madjarov2020, Martin2021, Pagano2022, Dlaska2022, Gonzalez2022}.
Moreover, Rydberg excited atoms provide a direct realization of interacting spin-$1/2$ models by encoding a spin between a ground state and Rydberg state \cite{Robicheaux2005, Weimer2008, Bernien2017, Choi2021}, or by directly encoding a spin in two Rydberg states \cite{DeLeseleuc2019, Signoles2021}, or off-resonant Rydberg dressing of ground state spins \cite{Henkel2010,Pupillo2010,Johnson2010,Jau2016, Zeiher2016, Hollerith2021}. 
Advances in optical trapping with tweezers \cite{schlosser2001sub, Kim2016, Endres2016, Barredo2016, Anderegg2019} and ground state cooling \cite{Lorenz2021,Ma2022,Kaufman2022} provide both a freely programmable geometric arrangement \cite{Barredo2018, Semeghini2021}, including coherent transport of atoms \cite{bluvstein2022quantum}, as well as scalability to  hundreds of particles \cite{Scholl2021, ebadi2021quantum}.
Experiments with circular Rydberg states \cite{Teixeira2020, Burgers2022,Muni2022} promise longer single particle lifetimes \cite{Kleppner1981}, with a future potential for scaling to even larger system sizes \cite{Nguyen2018,Cohen2021}.
We note, however, that up until now Rydberg platforms have mainly focused on quantum simulation of spin-$1/2$ systems.

\begin{figure*}[!t]
\center
\includegraphics[width=2.07\columnwidth]{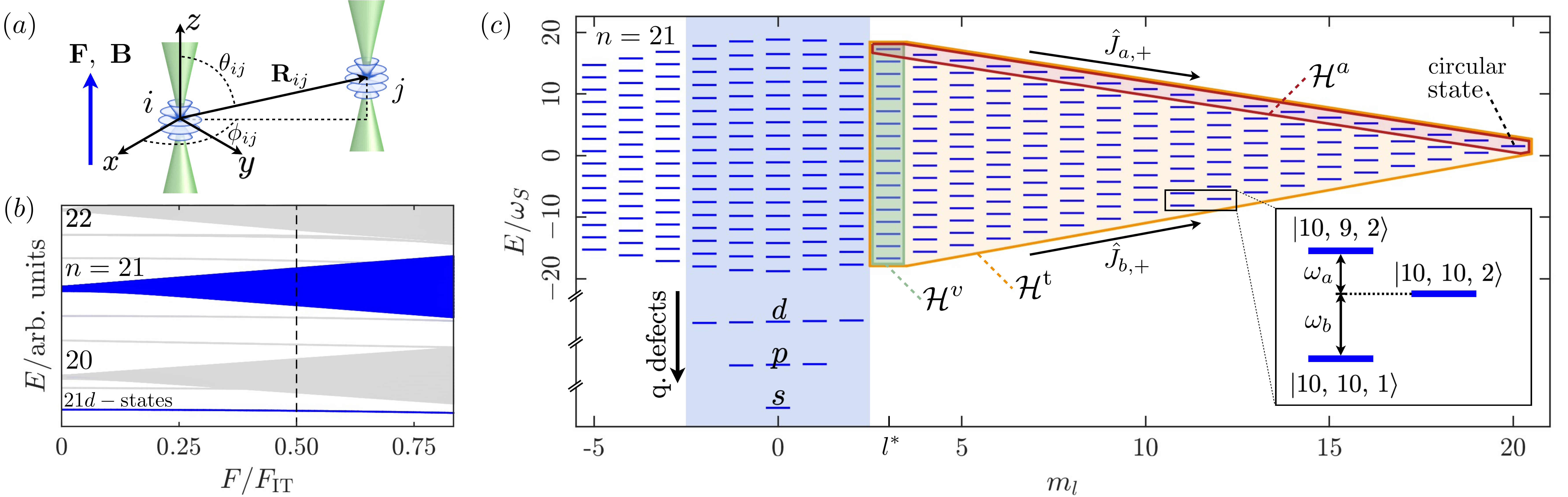}
\caption{SO(4)-symmetric $n$-manifold for quantum simulation. (a) Two tweezer trapped Rydberg atoms $i$ and $j$ in an external electric ($\mb F$) and magnetic ($\mb B$) field at a distance $\mb R_{ij}$ with relative orientation parametrized by the angles $\theta_{ij}$ and $\phi_{ij}$. 
(b) Numerically calculated Stark effect 
for \textsuperscript{87}Rb, where energies corresponding to $n=21$ are colored in blue. 
The large amount of discrete energy levels emerging from highly degenerate $F=0$ manifold makes the spectrum appear continuous. 
Notice that for non-hydrogenic atoms low orbital angular momentum states are shifted away from the Stark manifold ($d$-states are shown). 
(c) Energy eigenvalues for the $n=21$ manifold sorted with respect to the quantum number $m_l$ for a typical electric field value (indicated by the dashed line in (c) for $F = F_\mr{IT}/2$) $F_\mr{IT}>F>0$ and parallel magnetic field $B>0$, with $\omega_Z = 0.1\omega_S$, forming a rhombus shaped level structure. 
The manifold of states is described by two large angular momenta, where the application of raising operators $\hat J_{a,+}$ and $\hat J_{b,+}$ connects neighbouring states along the diagonal direction indicated by the corresponding arrow.
For non-hydrogenic atoms the highly regular spacing of the rhombus for $|m_l| < l^*$ (blue shaded area) is disturbed as low angular momentum states are shifted to lower energies due to quantum defects. 
For \textsuperscript{87}Rb the quantum defects 
are significantly nonzero for $l\lesssim3$ \cite{quantumdefectRb_spd, quantumdefectRb_f}, and for sufficiently large electric fields as considered here, only $s,p,d$-states are effectively missing and therefore we have $l^* = 3$.
The bordered and shaded areas indicate different manifolds of states, namely $\mathcal{H}^t$ (orange),  $\mathcal{H}^a$ (red) and $\mathcal{H}^v$ (green), which are used in this paper for quantum simulation purposes. 
The inset is a zoom of the level structure where the eigenstates and energy differences $\omega_a$ and $\omega_b$ are explicitly indicated.}
\label{fig:Fig1}
\end{figure*}

In contrast, we now investigate opportunities offered by the larger Rydberg $n$-manifold of states of Alkali, Alkaline Earth and Lanthanide atoms trapped in optical tweezers in the regime of a linear Stark and Zeeman effect, as illustrated in Fig.~\ref{fig:Fig1}. 
We use the SO(4) Lie algebra represented by two large angular momentum operators, which we denote by $\hat{\mb{J}}_a$ and $\hat{\mb{J}}_b$, to characterize the hydrogen-like high orbital angular momentum states and the action of the external electromagnetic fields.
Within this framework, we show how to encode and manipulate quantum states in these high-dimensional Rydberg manifolds. 
Moreover, this algebraic representation in terms of two angular momenta allows us to translate the dipole-dipole interactions between Rydberg atoms into generalized (large angular momentum) Heisenberg models, for which we discuss state preparation and readout schemes.

By exploiting the mapping of large angular momentum operators into conjugated (continuous) variables, we show that such Heisenberg models can be used to simulate Quantum Field Theories.
In particular, we illustrate the case of the sine-Gordon model and its massive extension, which is dual to 1+1D Quantum Electrodynamics, namely the Schwinger model.
Furthermore, we also discuss opportunities for quantum information processing offered by the Rydberg $n$-manifolds. 
We exemplify this for states near the circular level by providing a state-transfer protocol that can be employed to realize entangling gates for qudit-based quantum computation.

A timely opportunity to implement the results of our work is provided by recent Rydberg experiments with open inner shell atoms such as Er \cite{Trautmann2021}. 
First, the hydrogen-like high orbital angular momentum states can be directly accessed by laser light from low-lying atomic states due to the admixing of the core configurations with large angular momentum components.
Second, Alkaline Earth and Lanthanides offer the possibility to optically trap the atoms in Rydberg states via the valence core electrons and the corresponding ion core polarizability \cite{Mukherjee2011, Topcu2014, Teixeira2020, Wilson2022}. 

This article is structured as follows.
In Sec.~\ref{sec:atomicphysics}, we introduce the SO(4) formalism, the coupling to external fields, the form of dipole-dipole interactions and the role of quantum defects.
In Sec.~\ref{sec:manybodymodels}, we discuss how to encode and manipulate quantum information in the $n$-manifolds of Rydberg atoms, and we show the corresponding generalized Heisenberg models for Rydberg atoms in tweezer arrays.
In Sec.~\ref{sec:SGmodel}, we demonstrate how our approach can be used to simulate continuous variable systems, and in particular we discuss the paradigmatic sine-Gordon model and its massive extension.
In Sec. \ref{sec:ProspectQI}, we discuss a protocol to perform state transfer and entangling gates between two atoms within the manifold of states near the circular level. 
We conclude our work in  Sec.~\ref{sec:conclusions} with an outlook on future perspectives.

%%%%%%%%%%%%%%% Toolbox %%%%%%%%%%%%%%%%
\section{SO(4)-symmetric Rydberg manifolds}
\label{sec:atomicphysics}

In this work, we consider atoms trapped in tweezer arrays (e.g. Alkali, Earth-Alkaline and Lanthanide atoms) that are excited to hydrogen-like Rydberg states with principal quantum number $n$ in the regime of linear Stark and Zeeman effect. 
The corresponding manifold of $n^2$ levels displays a clean and regular structure originating from an underlying SO(4) symmetry described by two large angular momenta $\hat{\mb{J}}_a$ and $\hat{\mb{J}}_b$, as illustrated in Fig.~\ref{fig:Fig1}, which provide an efficient description to treat the single- and many-body problem. 
In this Section, we introduce the angular momentum structure of the Hydrogen problem, the resulting Rydberg state manipulation with electromagnetic fields and interactions, and discuss how these results translate to experimentally relevant multi-electron atoms.

\subsection{Algebraic solution of the Hydrogen problem}
\label{sec:algebraicsolution}

The SO(4) symmetry of the Hydrogen atom~\cite{Pauli1926, Biedenharn1984, Zee2016} allows for an elegant algebraic description of the Rydberg manifold of states with a fixed principal quantum number $n$. 
In particular we consider the non-relativistic Hydrogen Hamiltonian, \mbox{$\hat H_\mr{H} = {\hat{\mb{p}}^2}/{(2m_r)} - {e^2}/(4\pi\epsilon_0|\hat{\mb r}|)$}, where we ignore electronic spin degrees of freedom and $m_r$ is the reduced electron mass, $-e$ the electron charge, $\epsilon_0$ the vacuum permittivity and $\hat{\mb{p}}$ and $\hat{\mb r}$ are the momentum and position operators, respectively. 

A convenient approach to solve for the bound states of $\hat{H}_\mr{H}$ is to exploit the conservation of orbital angular momentum $\hatboldL$ and Runge-Lenz (RL) vector $\hat{\mb A}$ (see App.~\ref{sec:AppendixHydrogen} for details).
These two sets of conserved quantities generate the SO(4) symmetry of the Hydrogen Hamiltonian.
As the SO(4) group is doubly covered by SU(2)$\times$SU(2), we can construct two commuting angular momentum operators that read
\begin{align}
\hat{\mb J}_a = \frac{1}{2}\left( \hatboldL -\hat{\mb A}\right)~\mathrm{and}~ \hat{\mb J}_b = \frac{1}{2}\left( \hatboldL +\hat{\mb A}\right)\,,
\label{eq:JaJb}
\end{align}
and therefore obey $[\hat J_{a,i},\hat J_{a,j}] = i\epsilon_{ijk} \hat J_{a,k}$ and $[\hat J_{b,i},\hat J_{b,j}] = i\epsilon_{ijk} \hat J_{b,k}$, furthermore, the angular momentum raising and lowering operators are defined as $\hat J_{a(b),\pm} = \hat J_{a(b),x} \pm i \hat J_{a(b),y}$. 
Note, in this paper we use units where $\hbar = 1$.
Due to the orthogonality between the orbital angular momentum and the RL vector ($\hatboldL \!\cdot\! \hat{\mb A} = 0$), the lengths of the two angular momenta are constrained, $\hat{\mb J}_a^2 = \hat{\mb J}_b^2$.

Within this formalism, the Hamiltonian can be recast in the simple form $\hat H_\mathrm{H} = -\mr{Ry}/[2(\hat{\mb J}_a^2 +  \hat{\mb J}_b^2)+1]$, see App.~\ref{sec:AppendixHydrogen}, where $\mr{Ry} = e^2/(8\pi\epsilon_0a_0)$ is the Rydberg energy and $a_0 = 4\pi\epsilon_0/(e^2m_r)$ the Bohr radius. A natural basis for the hydrogen atom eigenstates is therefore given by states $\ket{J,m_a,m_b}$, where
\begin{align}
\label{eq:Eigenstates}
\hat{\mb{J}}_{a\,(b)}^2\ket{J,m_a,m_b} &= J(J+1)\ket{J,m_a,m_b}~\mr{and}\notag\\
\hat{J}_{a\,(b),z} \ket{J,m_a,m_b} &= m_{a\,(b)}\ket{J,m_a,m_b},
\end{align}
with $J = 0,1/2,1,\dots$ and $m_{a\,(b)}\in\{-J,\,-J+1,\,\dots,\,J\}$. 
The corresponding energies of the Hamiltonian take the form $E_n = -\mr{Ry}/(2J+1)^2 = -\mr{Ry}/n^2$ from which we can read off the relation between the principal quantum number $n$ and angular momentum length $J = (n-1)/2$. 
Therefore, a manifold of states with fixed $n$ hosts $n^2$ degenerate states $\ket{J,m_a,m_b}$, which can be labelled by the quantum numbers of two angular momenta.

A final result of the angular momentum formalism discussed in this Subsection is the form of the electric dipole operator $\hat{\bs{\mu}} = - e\hat{\mb r}$ whose transition matrix elements, within a single $n$ manifold, can be identified as 
\begin{align}
\hat{\bs{\mu}} = -3n\, e a_0\, \hat{\mb A}/2 = 3n\,e a_0\,\big(\hat{\mb J}_a - \hat{\mb J}_b\big)/2\,,
\label{eq:dip}
\end{align}
see Refs.~\cite{Pauli1926, Flamand1966, Becker1976, Valent2003} and App.~\ref{sec:AppendixHydrogen}. 
In the following, we will exploit this relation to analyse the angular momentum level structure in the presence of external electric fields and to express dipole-dipole interactions in terms of $\hat{\mb J}_a$ and $\hat{\mb J}_b$.

\subsection{Coupling to external fields}
\label{sec:externalfields}

The coupling of the electron to weak static external electric $\mb F$ and magnetic $\mb B$ fields lifts the degeneracy of states with fixed $n$, and is described by the Hamiltonian \begin{align}
\hat H_\mr{\mb{B},\mb{F}} = \mu_B \hatboldL\cdot\mb{B}-\hat{\bs{\mu}}\cdot\mb{F}\,,
\label{eq:FBHam}
\end{align}
where $\mu_B = e/(2m_r)$ is the Bohr magneton. 
For a manifold of hydrogen-like states with a specific $n$, the $\hatboldL$ and $\hat{\bs{\mu}}$ can be replaced by angular momentum operators according to Eq.~\eqref{eq:JaJb} and \eqref{eq:dip}, respectively.
The replacement of $\hat{\bs{\mu}}$ is valid below the Ingris-Teller limit $|\bs{F}|\!<\!F_\mr{IT} = 2\mr{Ry}/(3\,ea_0\,n^5)$, i.e. when couplings to adjacent $n' = n\pm 1$ manifolds are weak with respect to the gap $E_n-E_{n\pm 1}$ \cite{Demkov1970, Delande1988NewMethod, gallagher2006rydberg}.
Analogously, diamagnetic couplings are negligible in the limit \mbox{$|\mb{B}|< 2\mr{Ry}/(\mu_B\, n^4)$}~ \cite{Gay1983}.

In the following analysis, we focus on a single manifold of states with fixed $n$ and take the two fields to be parallel and oriented along the $z$-axis, $\mb F = F \mb e_z$ and $\mb B = B \mb e_z$, see Fig.~\ref{fig:Fig1}a, such that the projection $m_l = m_a + m_b$ of the orbital angular momentum $\hat L_z$ remains a good quantum number.
Within the limits discussed in the previous paragraph the Hamiltonian from  Eq.~\eqref{eq:FBHam} becomes 
\begin{align}
\hat H_n = -\omega_a\,\hat J_{a,z} + \omega_b\,\hat J_{b,z}\,,
\label{eq:static}
\end{align}
with $\omega_a = \omega_S - \omega_Z$ and $\omega_b = \omega_S + \omega_Z$, where $\omega_{S} = 3n\,e a_0\,F/2$ is the Stark splitting and $\omega_Z = \mu_BB$ is the Zeeman splitting (Larmor frequency).  
The linear Stark and Zeeman effect lifts the energy levels degeneracy, as shown in  Fig.~\ref{fig:Fig1}b as a function of the electric field strength for {}\textsuperscript{87}Rb.

The resulting spectrum, at fixed electric field, takes a regular spacing that forms a rhombic-shaped level structure \cite{born1925vorlesungen, Demkov1970, haroche2006exploring}, as shown in Fig.~\ref{fig:Fig1}c.
Deviations around $m_l=0$ of the regular spacing occur due to core corrections for non-hydrogenic atoms, which we discuss below in Subsec.~\ref{sec:generalization}.
For positive values of the electric field of a few $V/\mr{cm}$, and for magnetic fields of hundreds of Gauss, and principal quantum numbers in the range $30 \!<\! n \!<\! 70$, we have $\omega_Z<\omega_S$ such that $\omega_{a(b)}>0$. 
Within this parameter regime the absolute values of $\omega_a $ and $\omega_b$ can be of the order of $2\pi\times(10^2 - 10^3)\, \mathrm{MHz}$, which allows for direct coupling between the angular momentum states with microwave (MW) radiation.

Let us consider two circularly polarized MW fields whose electric field amplitudes are given by \mbox{$\mb{F}_a = F_a\mb e_-\exp(-i\omega_{a}^{\scriptscriptstyle\mr{MW}}t) + \mr{c.c.}$}, and \mbox{$\mb{F}_b = F_b\mb e_+\exp(-i\omega_{b}^{\scriptscriptstyle\mr{MW}}t) + \mr{c.c.}$}, with frequencies $\omega_{a(b)}^{\scriptscriptstyle\mr{MW}}$, electric field amplitudes $F_{a(b)}$ and polarizations \mbox{$\mb{e}_{\pm} = \mp(\mb{e}_{x} \pm i \mb{e}_{y})/\sqrt{2}$}. 
The electric field of the MWs couple to the dipole operator $\hat{\bs{\mu}}$, which for hydrogen-like states is represented by  Eq.~\eqref{eq:dip}. 
By combining the effect of the static and time-dependent fields and after going to the rotating frame, Eqs.~\eqref{eq:FBHam} and \eqref{eq:static} give rise to the most general Hamiltonian linear in angular momentum operators 
\begin{align}
\label{eq:MW}
\hat H_\mr{MW}=\sum_{\sigma=a,b}\left[-\Delta_{\sigma}\,\hat{J}_{\sigma,z} + \left(\Omega_\sigma\hat J_{\sigma,+} + \Omega^*_\sigma\hat J_{\sigma,-} \right)\right],
\end{align}
where we dropped fast oscillating terms and the rotating frame transformation is defined by $\hat R =  \exp[-i(\omega_a^{\scriptscriptstyle\mr{MW}}\hat{J}_{a,z}-\omega_{b}^{\scriptscriptstyle\mr{MW}}\hat{J}_{b,z})t]$.
Here, $\Delta_{a} =  \omega_a-\omega_{a}^{\scriptscriptstyle\mr{MW}}$ ($\Delta_{b} =  \omega_{b}^{\scriptscriptstyle\mr{MW}}-\omega_{b}$) and $\Omega_{a(b)} = 3n\,ea_0\, F_{a(b)}/(2\sqrt{2})$ are independently tunable  detunings and complex Rabi frequencies, respectively.
We note that such MW control has already been proposed \cite{Hulet1983} and demonstrated in recent experiments \cite{Signoles2017}. 
Deviations from perfectly circularly polarized MW fields can give rise to unwanted transitions, which can be suppressed by tuning the level spacing imbalance \mbox{$\omega_a-\omega_b$} via the magnetic field.

\subsection{Dipole-dipole interactions}

We now turn our attention to the role and form of interactions.
Let us consider a pair of atoms $i$ and $j$, with a relative position $\mb R_{ij} = R_{ij}\, \mb{e}_{ij}$, here $R_{ij}$ is the inter-atomic separation and \mbox{$\mb{e}_{ij} = (\cos\phi_{ij}\,\sin\theta_{ij}, \sin\phi_{ij} \sin\theta_{ij},\cos\theta_{ij})^\mr{T}$} is the corresponding unit vector, where $\theta_{ij}$ and $\phi_{ij}$ are the azimuthal and polar angle, respectively, see Fig.~\ref{fig:Fig1}a. 
For distances of a few to tens of micrometers, \emph{i.e.} relevant for optical tweezer experiments, the interaction between two atoms is dominated by the electric~\footnote{Magnetic dipole-dipole interactions are $\alpha^2/(9n^2)$ times smaller than electric dipole-dipole interactions , where $\alpha$ is the fine structure constant, and therefore we neglect them.} dipole-dipole coupling \cite{Buehler1951, Weber2017}  
\begin{align*}
   \hat H^{ij}_{\mathrm{dd}} = \frac{1}{4\pi\epsilon_0 R_{ij}^3} \left[\hat{ \bs{\mu}}^{(i)}\cdot \hat{\bs{\mu}}^{(j)} - 3 (\hat{\bs{\mu}}^{(i)}\cdot \mb{e}_{ij}) (\hat{\bs{\mu}}^{(j)}\cdot \mb{e}_{ij})  \right],  
\end{align*}
where $\hat{\bs{\mu}}^{(i)}$ and $\hat{\bs{\mu}}^{(j)}$ are the dipole operators of atom $i$ and $j$, respectively. 

We consider the situation where the two atoms $i$ and $j$ are initially prepared in a Rydberg $n$-manifold, and dipole-dipole interactions are weak as compared to the energy gap to adjacent $n' = n\pm 1$-manifolds, such that no transitions to other $n'$-manifolds occur. 
Within these criteria, the dipole operator can be expressed in terms of angular momentum operators through the application of Eq.~(\ref{eq:dip}), and the interaction Hamiltonian can be compactly written as
\begin{align}
\hat H^{ij}_{\mathrm{dd}} = V_{ij} \Big[&\hat{\mb J}^{(i)}_a\cdot \hat{\mb J}^{(j)}_a - 3 (\hat{\mb J}^{(i)}_a\cdot \mb{e}_{ij}) (\hat{\mb J}^{(j)}_a\cdot \mb{e}_{ij})  \notag\\
+&\hat{\mb J}^{(i)}_b\cdot \hat{\mb J}^{(j)}_b - 3 (\hat{\mb J}^{(i)}_b\cdot \mb{e}_{ij}) (\hat{\mb J}^{(j)}_b\cdot \mb{e}_{ij}) \notag\\
-&\hat{\mb J}^{(i)}_a\cdot \hat{\mb J}^{(j)}_b + 3 (\hat{\mb J}^{(i)}_a\cdot \mb{e}_{ij}) (\hat{\mb J}^{(j)}_b\cdot \mb{e}_{ij}) \notag\\
- &\hat{\mb J}^{(i)}_b\cdot \hat{\mb J}^{(j)}_a + 3 (\hat{\mb J}^{(i)}_b\cdot \mb{e}_{ij}) (\hat{\mb J}^{(j)}_a\cdot \mb{e}_{ij})\Big],
\label{eq:AMgeneral}
\end{align}
with $V_{ij} = (3n\,e a_0)^2/(16\pi\epsilon_0R_{ij}^3)$. 
Similarly to other dipolar systems \cite{Baranov2012}, the spatial anisotropy ($\theta_{ij}$ and $\phi_{ij}$) of interactions and the geometrical arrangement of the atoms can be exploited to tune the dipole-dipole interactions. 
Furthermore, time-dependent methods \cite{Pfau2002, Choi2020, DiLiberto2021}
can also be applied to control and design the interaction Hamiltonian, as recently experimentally demonstrated for magnetic dipolar atoms \cite{Lev2020} and for Rydberg atoms \cite{Geier2021,Browaeys2022}.
This will therefore allow us to realize and tune a whole class of interaction Hamiltonians.

\subsection{Nonhydrogenic atoms and the electron spin}
\label{sec:generalization}

The Rydberg manifolds of states with $n\gg 1$ of experimentally relevant Alkali, Alkaline Earth and Lanthanide atoms follow, up to core corrections, the SO(4) symmetric hydrogenic results presented in Subsec.~\ref{sec:algebraicsolution}. 
In particular, only a few low orbital angular momentum states, which are typically $l=s,p,d(,f)$ states, penetrate the core region and are therefore energetically shifted away from the $n$ manifold under consideration.
Their energy shift is described by quantum defect theory \cite{Seaton1983, Aymar1996} and is given by $E_{n,l} = -\mr{Ry}/(n-\delta_l)^2$, where $\delta_l$ is the quantum defect \footnote{For atomic species with more than one valence electron, like Alkaline Earth or Lanthanide atoms, multichannel quantum defect theory has to be applied in order to calculate the spectrum \cite{Vaillant2014}. 
For these species, we consider a Rydberg manifold $n$ where interactions between different Rydberg series, corresponding to different core configurations, are negligible.}.

Including the quantum defects, the Rydberg Hamiltonian from Eq.~\eqref{eq:static} for a given $n$ manifold becomes \footnote{States affected by quantum defects acquire an additional phase shift proportional to $\delta_l$ with respect to their hydrogenic counterpart and experience a modified core potential including spin-orbit effects. 
Therefore, the wave functions are different from hydrogen ones and, hence, dipole transition matrix elements to these states are not captured by Eq.~\eqref{eq:dip}. However, this modification does not alter our results as only states unaffected by quantum defects are considered for quantum simulation applications.}
\begin{align}
\hat H_n^\mr{QD} = \hat H_n +\!\!\sum_{\substack{|m_l|\leq l \\~~ l<l^*}}\ket{n,l,m_l} \bra{n,l,m_l} (E_{n,l} - E_n),
\end{align}
where the threshold value $l^*$ is dictated by the magnitude of the quantum defects, which is different for each atomic species \cite{Fano1976}.
For example, for the case of Rb, and sufficiently strong electric fields, $l^* = 3$ while for Er  typically $l^* = 4$.
In contrast to the hydrogenic case, the energies of $\hat H_n^\mr{QD}$ do not form a perfectly regular rhombus structure anymore because the inner region \mbox{$|m_l|\!<\!l^*$} presents an irregular spacing \cite{Fabre1984, Delande1988NewMethod}. 
However, for $n\gg 1$ the large majority of states follow the hydrogenic description, as shown in Fig.~\ref{fig:Fig1}c for the $n = 21$ manifold. 

The treatment of non-relativistic Hydrogen has thus far ignored the effect of spin orbit coupling, which in fact should be taken into account for realistic atoms.
Nevertheless, spin orbit coupling primarily only affects the low $l$ states which are shifted away from  a Rydberg manifold $n$ due to quantum defects. The spin orbit coupling of high $l$ states instead, is negligible for sufficiently large magnetic fields, due to a strong-field spin Zeeman (or Paschen-Back) effect \cite{sobelman2012atomic}.
Hence, the rhombus manifold, which consists of only states with high $l$, decouples from the electron spin an thus comes in two copies, $\ket{J, m_a, m_b}\otimes\ket{s,m_s}$ with the electron spin state $\ket{s,m_s} = \ket{1/2, \pm1/2}$, energetically resolvable through the spin Zeeman splitting. 
This provides in principle a further degree of freedom that  doubles the available state space, but in the remainder of this paper we will focus on a single spin state $\ket{s,m_s} = \ket{1/2, 1/2}$.

Let us finally comment on possible decoherence mechanisms of the hydrogen-like Rydberg states of a given $n$-manifold. 
First, radiative decay due to spontaneous emission and black body radiation is a well known source of decoherence in Rydberg experiments \cite{Low2012, Goldschmidt2016, Festa2022}. 
However, similar to experiments with circular Rydberg states~\cite{Nguyen2018,Cantat-Moltrecht2020}, black body radiation can be suppressed in  a cryogenic environment, while at the same time spontaneous emission is strongly reduced for all states with $m_l > l^*$ since the primary decay channels to energetically low-lying states are dipole-forbidden (for further details see App.~\ref{Appendix:SpontaneousEmission}).
Radiative decay rates can therefore safely be assumed to be well below all other relevant energy scales in this work. 

Further sources of decoherence can for instance originate from fluctuations in electric and magnetic fields, which directly affect the energy levels $\omega_{a(b)}$, and therefore enter as dephasing. 
Experimentally demonstrated electric field stabilizations on the level of tens of $\mu V/\mr{cm}$ \cite{Facon2016} and magnetic field stabilization of tens of $\mu$G \cite{Hesse2021} should however be sufficient to render these decoherence effects negligible. 
Another source of decoherence  for tweezer trapped Rydberg atoms is motional dephasing, due to the trapping potential being dependent on the internal state.
Trapping of the core-electron, as is possible for Alkaline-Earth and Lanthanide atoms, would remove this problem by providing state insensitive traps \cite{Topcu2014, Mukherjee2011, Wilson2022}, see App.~\ref{Appendix:MotionalDecoherence}.
Furthermore, atoms internally prepared in the Rydberg state experience (state dependent) dipole-dipole forces, which leads to heating of the initially laser-cooled \cite{Kaufman2022,Ma2022} motional state.
These heating rates are negligible for the inter-atom distances of few tens of $\mu$m and for the time scales considered in this manuscript, see App.~\ref{Appendix:MotionalDecoherence}. 

In summary, Rydberg states with a fixed principal quantum number $n$ of experimentally relevant atomic species provide a natural setting for quantum simulation and quantum information processing with a large internal state space.
Except for very few states that are affected by quantum defects, the Rydberg $n$-manifold inherits the physics of the hydrogen states. 
This includes the single-particle couplings to external fields, see Eq.~\eqref{eq:MW}, and two-body dipole-dipole interactions, see Eq.~\eqref{eq:AMgeneral}, giving rise to generalized Heisenberg models for arrays of many atoms. 
In the next Section, we consider different subsets of the $n$-manifold and illustrate several concrete examples of many-body Hamiltonians that can be naturally implemented with this platform.

\section{Engineering of quantum many-body models}
\label{sec:manybodymodels}

In this Section, we consider different subsets of the $n$-manifold in order to engineer specific examples of many-body Hamiltonians for the hydrogen-like states, thus avoiding the quantum defect modified region.
More specifically, we are selecting two manifolds that are closed under dipole-dipole interactions for quantum simulation and quantum information processing and one manifold that can be employed for state preparation and measurement.

\emph{Triangular manifold}. The first and largest submanifold closed under dipole-dipole interactions that we consider is the triangular manifold defined for a fixed $n$ by 
\begin{align}
\label{eq:tmanif}
\mathcal H^t  = \big\{\ket{\psi^t_{m_a,m_b}}\big\},\, \mr{with}\,\, |\psi^t_{m_a,m_b}\rangle \equiv |J, m_a, m_b \rangle\,,
\end{align}
and $m_l\geq l^*$. This defines the set of all states unaffected by quantum defects on the right side of the rhombus and highlighted by the orange triangle in Fig.~\ref{fig:Fig1}c.
In addition to dipole-dipole interactions and single-particle control, we will demonstrate how to design nonlinear terms in the angular momentum operators with ponderomotive manipulation techniques that can be used for different kinds of state manipulation and engineering schemes.
Later in this work, these nonlinear terms will be instrumental for the analog simulation of Quantum Field Theories.
Furthermore, near the circular level we derive an effective description by using the Holstein-Primakoff transformation, which enables us to perform a state-transfer operation between pairs of atoms as a building block for qudit-based quantum computing.

\emph{Edge manifold}. The second manifold closed under dipole-dipole interactions that we consider is a subset of the triangular manifold $\mathcal H^t$ and is defined by the states with maximally-polarized angular momentum $\hat{\mathbf J}_b$, namely
\begin{align}
\label{eq:amanif}
\mathcal H^a = \big\{\ket{\psi^{a}_{m_a}} \big\},\, \mr{with}\,\, |\psi^{a}_{m_a}\rangle \equiv |J, m_a, m_b \!\!=\!\! J \rangle \,,
\end{align}
and $m_a+m_b\geq l^*$, which is highlighted in red in Fig.~\ref{fig:Fig1}c. 
In addition to the control offered by the triangular manifold, here we can also engineer additional nonlinear squeezing terms by off-resonantly MW coupling between different $n' > n$ manifolds.
This additional ingredient will provide the kinetic term required for the simulation of Quantum Field Theories.

\emph{Vertical manifold}. The last manifold considered in our discussion is the vertical manifold defined by
\begin{align}
\label{eq:vmanif}
\mathcal H^v \! = \! \big\{\ket{\psi^{v}_{m_a}} \big\}\,, \mr{with}\,\, |\psi^{v}_{m_a}\rangle \!\equiv\! |J, m_a, m_b \!=\! l^*\!-\!m_a \rangle ,
\end{align}
and fixed $m_l \!=\! l^*$, which defines a set of states immediately next to the quantum defect modified region highlighted in green in Fig.~\ref{fig:Fig1}c.
Although this set of states is not closed under dipole-dipole interactions, its unique properties can be exploited for state preparation and readout for the $\mathcal H^a$ manifold.

\begin{figure}[!t]
\center
\includegraphics[width=1\columnwidth]{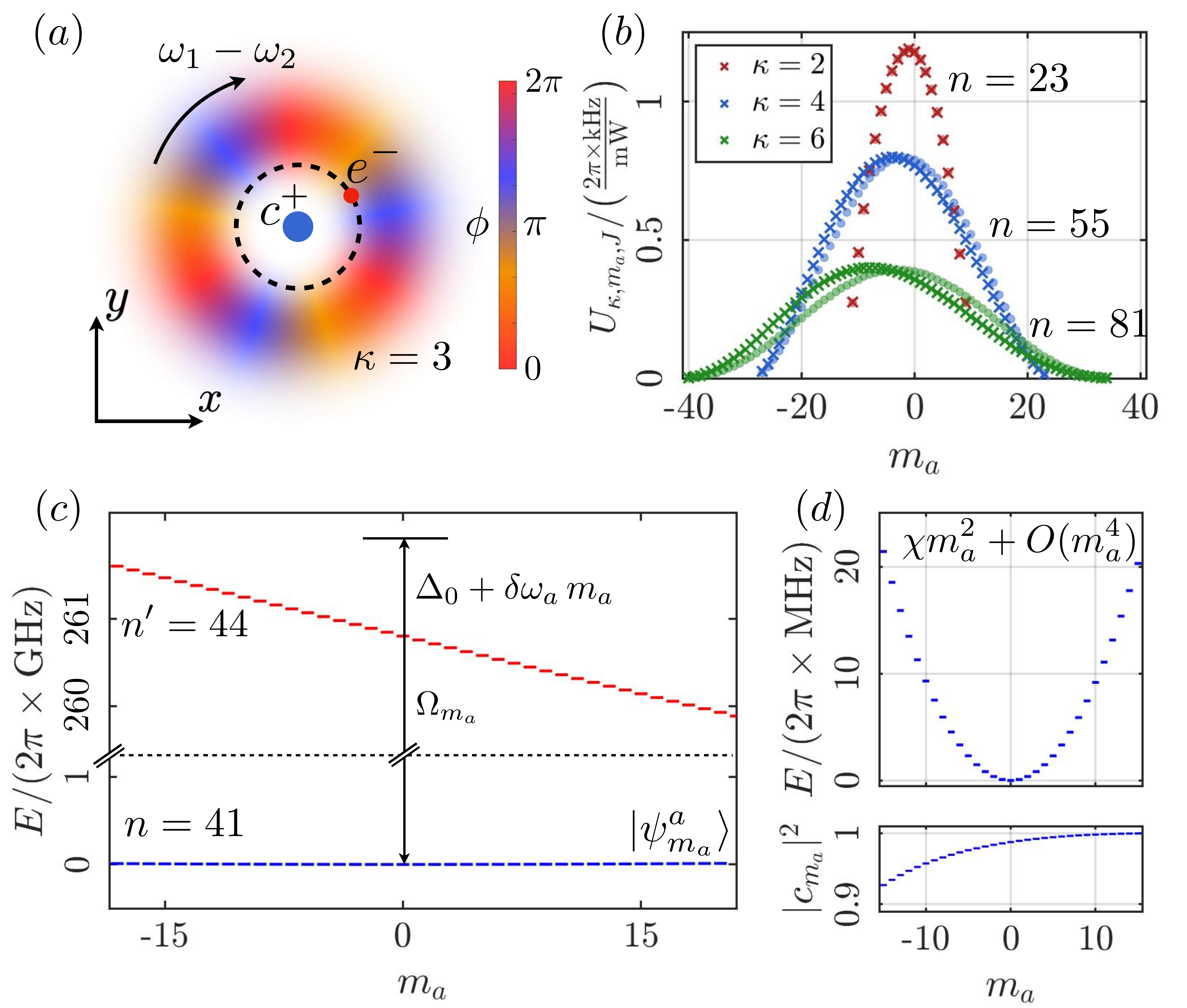}
\caption{Nonlinear control: 
(a) Illustration of the time-dependent part of the ponderomotive potential \mbox{$U_\kappa(x,y,z=0) e^{-i\kappa\phi}$}, see Eq.~\eqref{eq:PMphase} in App.~\ref{subsec:ponderomotive}. 
The color intensity is proportional to the potential strength $U_\kappa$ and the color code indicates the phase pattern $\phi$.
The blue dot represents the ionic core $c^+$, whereas the dashed circle represents the spatial extent of a typical valence electron $e^-$ Rydberg wavefunction as compared to the ponderomotive potential extent. 
(b) Comparison of numerically computed transition matrix elements
\mbox{$U_{\kappa,m_a,m_b}\equiv\bra{\psi^t_{m_a+\kappa,m_b}} U_\kappa(\mb{r}) e^{-i\kappa\phi} \ket{\psi^t_{m_a,m_b}}$}
(crosses) with the expression $\lambda_{\kappa} \bra{\psi^a_{m_a+\kappa,m_b}} (\hat J_{a,+})^\kappa \ket{\psi^a_{m_a,m_b}}$ (circles) for different values of $n$, $\kappa$ and $m_b = J$. 
The parameter $\lambda_\kappa$ is extracted from a least-square fit.
The coupling strengths are normalized to the sum of the two laser beams powers with wavelength $\lambda = 1300\,\mr{nm}$ and beam waist $\lambda/2$.  
(c) Illustration of the off-resonant MW coupling between the $\mathcal H^a$ manifolds with $n = 41$ (blue) and $n' = 44$ (red). 
The static electric field is chosen to be $F=F_\mr{IT}/2$ and for illustrative purposes we choose the magnetic field such that $\omega_a(n) = 0$. 
(d) The upper panel displays the AC stark shift of the states $\ket{\psi^a_{m_a}}$ which is quadratic in the vicinity of $m_a=0$. 
The lower panel monitors the purity of the dressed eigenstates $|c_{m_a}|^2$, which decreases for $m_a\rightarrow -J$ as the MW detuning is $m_a$ dependent as shown in (c).
For both panels the electric field is chosen as in (c), the magnetic field $B>0$ is chosen such that $\omega_Z = \omega_S/2$, whereas the other parameters are $\Omega_{m_a = 0} =2\pi\times  319\,\mr{MHz}$, $\Delta_0 = 2\pi\times 1.4\,\mr{GHz}$  leading to $\chi = 2\pi\times 92\,\mathrm{kHz}$.}
\label{fig:MWPO}
\end{figure}

\subsection{Triangular manifold \texorpdfstring{$\mathcal H^t$}{Ht}}
\label{subsec:triangular}

The triangular manifold $\mathcal H^t$ is defined in Eq.~\eqref{eq:tmanif} and is highlighted in orange in Fig.~\ref{fig:Fig1}c. 
Here below we discuss a specific example of many-body Hamiltonian, the ponderomotive manipulation technique to engineer a class of nonlinear terms and we provide an effective theory for the states near the circular level.

\subsubsection{Model Hamiltonian}

To be specific, we consider the case where all atoms are placed in a plane perpendicular to the static electric and magnetic fields, $\theta_{ij} = \pi/2$, and focus on the regime \mbox{$|\omega_{a(b)}| \gg J^2 V_{ij}$}, namely when the Stark splitting is much larger than the typical interaction energy scales. 
By considering the single-particle and many-body terms introduced in the previous Section in Eqs.~\eqref{eq:MW} and \eqref{eq:AMgeneral}, with the addition of the ponderomotive drive that we detail in the next Subsection, the many-body Hamiltonian takes the form
\begin{align}
    \label{eq:abHeisenberg}
    &\hat H^t = \sum_i\,\left(\hat H^{(i)}_\mr{MW} + \hat H^{(i)}_\mr{P}\right)\notag\\
    &~+\frac{1}{2}\sum_{i\neq j} V_{ij}\bigg[\left(\hat J_{a,z}^{(i)}- \hat J_{b,z}^{(i)}\right)\left(\hat J_{a,z}^{(j)}-\hat J_{b,z}^{(j)}\right)\\
    &~-\frac{1}{4}\sum_\sigma\,\left(\hat J^{(i)}_{\sigma,+}\hat J^{(j)}_{\sigma,-} \!+\! \mr{H.c.} \right) \!+\! \frac{3}{2} \left(e^{i2\phi_{ij}} \hat  J^{(i)}_{a,+} \hat  J^{(j)}_{b,+} \!+\! \mr{H.c.}\right)\bigg],\notag
\end{align}
where we transformed to the rotating frame defined below Eq.~\eqref{eq:MW} under the resonant condition $\Delta_a \!+\! \Delta_b = \omega_a \!-\! \omega_b$. 
The term $\hat H^{(i)}_\mr{P}$ is the ponderomotive coupling for the $i$-th atom defined as 
\begin{align}
\label{eq:ponderomotive}
\hat H_\mr{P}\! =\sum_{\kappa_a,\kappa_b} \lambda_{\kappa_a, \kappa_b}  
(\hat J_{a,+})^{\kappa_a} (\hat J_{b,+})^{\kappa_b} +\mr{H.c.}\,,
\end{align}
where $\lambda_{\kappa_a, \kappa_b}$ denotes complex coupling strengths. 
The ponderomotive coupling in Eq.~\eqref{eq:ponderomotive} corresponds to a transfer of orbital angular momentum by $\kappa = \kappa_a + \kappa_b$, where  $\kappa_{a(b)}$ can take on negative and positive values (thus with the identification $(\hat J_{a(b),+})^{-1} \rightarrow \hat J_{a(b),-}$). 
For the special case $\kappa_b=0$ implying $\kappa = \kappa_a$, the ponderomotive Hamiltonian $\hat H_{\mr{P}}$ reduces to $\lambda_\kappa(\hat J_{a,+})^\kappa + \lambda_\kappa^*(\hat J_{a,-})^\kappa$, where we use the simplified notation $ \lambda_\kappa \equiv \lambda_{\kappa,0}$.
In this case, this coupling Hamiltonian can be used, for instance, to generate a two-axis twisting term \cite{kitagawa1993squeezed}, \emph{i.e.} \mbox{$i(\hat J_{a,+})^2 -i(\hat J_{a,-})^2$} for $\kappa = 2$.

\subsubsection{Ponderomotive manipulation}
\label{subsubsec:Pomderomotive}

In order to realize the nonlinear coupling in Eq.~\eqref{eq:ponderomotive}, we employ the ponderomotive manipulation techniques \cite{Dutta2000,Knuffman2007, cardman2020circularizing, Cohen2021}.  
The core idea is to interfere two co-propagating (along the $z$-direction) hollow Laguerre-Gauss laser beams with different frequencies $\omega_1$ and $\omega_2$ and non-zero orbital angular momentum. 
The corresponding interference pattern forms a ponderomotive potential $U_\kappa(\mb{r})e^{-i\kappa \phi}e^{-i\delta\omega t}+\mathrm{H.c.}\,$, rotating at the beating frequency $\delta\omega =\omega_1-\omega_2$, which can be arranged to be in the MW frequency domain to match the resonance condition $\delta\omega = -\kappa_a\, \omega_a + \kappa_b\, \omega_b$. 
When centered at the atomic position, this spatially-rotating ponderomotive potential transfers $\kappa=\kappa_a + \kappa_b$ quanta of orbital angular momentum, see Fig.~\ref{fig:MWPO}a. 

For simplicity, we focus here on the case $\kappa_b \!=\! 0$ and therefore $\kappa \!=\! \kappa_a$. 
The transition matrix elements of the ponderomotive coupling \mbox{$U_{\kappa,m_a,m_b} \equiv \bra{\psi^t_{m_a+\kappa,m_b}}U_\kappa(\mb{r})e^{-i\kappa \phi}\ket{\psi^t_{m_a,m_b}}$} are shown in Fig.~\ref{fig:MWPO}b for $\kappa = 2,\, 4,\, 6$, $m_b = J$ and different values of $n$. 
As discussed in App.~\ref{subsec:ponderomotive}, when the beam waist is large compared to the Rydberg orbital radius, the leading order expression of these matrix elements is
\mbox{$U_{\kappa,m_a,m_b} =\lambda_\kappa \bra{\psi^t_{m_a+\kappa,m_b}}(\hat J_{a,+})^\kappa \ket{\psi^t_{m_a,m_b}}$}. 
For simplicity, we extract the coefficient $\lambda_\kappa$ by a least square fit and show the accuracy of this procedure in Fig.~\ref{fig:MWPO}b.
The degree of accuracy can be controlled by increasing the laser beam waist, which requires more laser power to maintain the same strength for the transition amplitudes. 
The parameter $\lambda_\kappa$ can be complex and is controlled by the relative phase of the Laguerre-Gauss beams.
Furthermore, the coupling matrix elements $U_{\kappa,m_a}$ scale as $n^{2\kappa}$ for $|m_a|\ll J$, and are in the range $2\pi\times(10^2-10^3)\,\mr{kHz}$ for hundreds of mW of laser power.
Further details of these calculations and precise beam parameters are discussed in App.~\ref{subsec:ponderomotive}.

We point out that for fixed laser power and $n$ achievable coupling strength decrease as $|\kappa_a| + |\kappa_b|$ is increased. 
For $|\kappa_a| + |\kappa_b| \lesssim 6$ and a combined laser power of hundred mW per site give rise to to coupling strength of hundreds of $2\pi\times$kHz, see Fig.~\ref{fig:MWPO}b.
Moreover, notice that the expression \eqref{eq:ponderomotive} also contains the linear expression \eqref{eq:MW} obtained via a MW driving as a special case. 
However, as the ponderomotive coupling is driven by laser light, it directly provides a way to locally engineer these couplings, i.e. at the level of a single Rydberg atom, which can be exploited for digital quantum simulation purposes.

\subsubsection{Effective theory near the circular state}
\label{subsubsec:NearCircularState}

A model Hamiltonian similar to Eq.~\eqref{eq:abHeisenberg} can also be realized with coupled atomic ensembles \cite{Hammerer2010}, where a single large angular momentum of length $J$ is constructed out of $2J$ two-level atoms. 
In our case two large angular momenta, both of length $J$, are encoded in a single atom with principal quantum number $n=2J+1$, which can be manipulated with external fields enabling, for example, the preparation of squeezed states, see Eq.~\eqref{eq:ponderomotive}.
While atomic ensembles are usually coupled via atom-light interactions, here pairs of angular momenta are naturally coupled by dipole-dipole forces. 

Analogous to atomic ensembles, where states close to the maximally stretched state are described by continuous variables \cite{Hammerer2010}, we now focus on the Hamiltonian \eqref{eq:abHeisenberg} on a subspace of states close to the maximally stretched (circular) level $\ket{\psi^t_{J,J}}$. 
We thus perform a Holstein-Primakoff transformation \mbox{$\hat J_{-,\sigma}^{(i)} = \sqrt{2J}\sqrt{1-\hat n_{i\sigma}/(2J)}\,\hat c_{i\sigma}$} and $\hat J_{\sigma,z}^{(i)} = 2(J-\hat n_{i\sigma})$, where $\hat c^{}_{i\sigma}$ ($\hat c^\dagger_{i\sigma}$) are standard bosonic annihilation (creation) operators satisfying the commutation relations $[\hat c^{}_{i\sigma},\hat c^\dagger_{j\sigma'}]=\delta_{\sigma\sigma'}\delta_{ij}$ and $\hat n_{i\sigma} = \hat c^\dagger_{i\sigma} \hat c^{}_{i\sigma}$, with $\sigma = a,b$ playing the role of pseudo-spin.

In this framework, the circular level serves as the vacuum for both bosonic modes, $\ket{0,0}\equiv\ket{\psi^t_{J,J}}$, and the Hamiltonian (without MW and ponderomotive couplings) takes the form of a generalized spinful Bose-Hubbard model
\begin{align}
\label{eq:HPHamiltonian}
    &\hat H^t_{\mr{HP}} \!=\!\! \sum_{\sigma=a,b}\bigg[\!-\!\sum_i\,\Delta_\sigma\,\hat n_{i\sigma} \!-\! \frac{1}{2}\sum_{i\neq j} h_{ij}\, \bigg(\hat c_{i\sigma}^\dagger \hat c_{j\sigma}^{}\!+\!\mr{H.c.}\bigg)\bigg]\\
    &~+ \frac{1}{2}\sum_{i\neq j} \left[\left(w_{ij}\hat c^\dagger_{ia}\hat c^\dagger_{jb} +\mr{H.c.}\right)+U_{ij}(\hat n_{ia}\!-\!\hat n_{ib})(\hat n_{ja}\!-\!\hat n_{jb})\right],\notag
\end{align}
which is valid up to corrections $O(1/J)$.
Here $\Delta_\sigma$ are onsite energies,  $h_{ij} = JV_{ij}/2$ describes intra-species hopping, $w_{ij} = 3J\exp[i2\phi_{ij}]$ pair gain/loss processes, and $U_{ij} = 2V_{ij}$ density-density interactions.
In contrast to standard Bose-Hubbard models the pair gain/loss term $w_{ij}$ violates total particle number conservation and can be used to generate parametric squeezing.
Further below in Sec.~\ref{sec:ProspectQI} we outline another application of  Eq.~\eqref{eq:HPHamiltonian}, where we employ the hopping term $h_{ij}$ for quantum state transfer and an entangling gate between a pair of atoms.

The physics of Eq.~\eqref{eq:HPHamiltonian} can be explored starting from, for example, the vacuum state $\ket{\psi^t_{J,J}}$, which can be initialized through well established circularization methods, \emph{i.e.} via the adiabatic rapid passage method \cite{Hulet1983,Liang1986,Nussenzveig1993,cheng1994production}, the crossed electric and mangnetic fields method \cite{Delande1988NewMethod,Morgan2018}, multi MW photon transfer \cite{Signoles2017, Larrouy2020} or direct ponderomotive coupling \cite{cardman2020circularizing}. 
Another possible scheme to prepare more general product states is outlined in Sec.~\ref{subsec:vertical}, where we describe a generalization of the adiabatic rapid passage protocol used to prepare circular levels.
Readout can be performed by applying energy-selective MW pulses that transfer the population of individual levels to different $n'$-manifolds, which are subsequently resolved by an electric field ionization measurement \cite{Signoles2017}.

\subsection{Edge manifold \texorpdfstring{$\mathcal H^a$}{Ha}}
\label{subsec:edge}

The edge manifold $\mathcal H^a$ is defined in Eq.~\eqref{eq:amanif} by having $\hat{\mb J}_b$ maximally polarized, as highlighted in Fig.~\ref{fig:Fig1}c. 
Notice that we could define an analogous manifold $\mathcal H^b$ by considering the states with $\hat{\mb J}_a$ maximally polarized.
Choosing the static external electric and magnetic fields such that $|\omega_{a(b)}|,|\omega_a \pm\omega_b| \gg J^2 V_{ij}$, this manifold is closed under dipole-dipole interaction processes governed by $\hat H^{ij}_\mr{dd}$.

Since $\mathcal H^a$ is a subset of $\mathcal H^t$, it inherits many of the properties discussed before, such as the ponderomotive control.
Additionally, we can also construct a nonlinear squeezing term 
\begin{equation}
    \hat H_{\mr{SQ}} = \chi (\hat J_{a,z})^2\,,
    \label{eq:Squeezing}
\end{equation}
where $\chi$ is the squeezing strength. This can be generated by off-resonantly coupling the $\mathcal H^a$ manifold with principal quantum number $n$ to another $\mathcal H^a$ manifold with $n'>n$, through MW radiation.

In the blue detuned regime, see Fig.~\ref{fig:MWPO}c, and $z$-polarized MW fields, the system can be treated as a collection of independent two-level systems, one for each $m_a$.
As shown in Fig.~\ref{fig:MWPO}c, the detuning $\Delta_{m_a} = \Delta_0 + \delta\omega_a m_a$ of each two-level system varies with $m_a$, where $\Delta_0$ is an overall detuning and $\delta\omega_a \!=\! \omega_a(n') \!-\! \omega_a(n)$ is the differential level shift.
Moreover, the coupling strength $\Omega_{m_a}$ depends smoothly on $m_a$ through the dipole transition matrix elements.
In the regime $\Delta_{m_a} \!\gg\! \Omega_{m_a}$ the coupling leads to an AC Stark shift $\Omega_{m_a}^2/(4\Delta_{m_a})$ that can be expanded in a power series of $m_a$ when $\Delta_0\!\gg\! \delta\omega_a m_a$, namely $\Omega_{m_a}^2/(4\Delta_{m_a}) =\chi m_a^2+ O(m_a^4)$, where the constant and linear terms are absorbed in $\Delta_0$ and $\delta\omega_a$, respectively. 
The (quadratic) leading term of this expansion therefore gives rise to the anticipated squeezing term, as shown in the upper panel of Fig.~\ref{fig:MWPO}d.
Cubic terms can be canceled exactly by carefully picking $\Delta_0$ and $\Omega_{m_a}$, as discussed in App.~\ref{Appendix:OffresonantMW}.

For moderate values of the MW intensity, $\chi$ can be on the order of $2\pi\times 100\,\mathrm{kHz}$, thus giving rise to level shifts as large as $2\pi\times 10\,\mathrm{MHz}$ for $|m_a|\sim J$.
For such values of the level shifts, a small admixture of states from the $n'$ manifold is present, see the lower panel of Fig.~\ref{fig:MWPO}d. 
Moreover, imperfect MW polarization can lead to unwanted two-photon Raman processes, which can be made off-resonant by tuning $\omega_a \gg \Omega_{m_a}^2/\Delta_{m_a}$.

By taking into account all the terms discussed so far, the Hamiltonian for the $\mathcal{H}^a$ manifold takes the form
\begin{align}
\hat{H}^a &\!=\! \sum_i\, \left\{ \!-\!\Delta_a \hat{J}_{a,z}^{(i)} \!+\! \chi \big(\hat{J}_{a,z}^{(i)}\big)^2 \!+\! \sum_\kappa \left[\lambda_{\kappa} \big(\hat{J}_{a,+}^{(i)}\big)^\kappa \!+\! \mr{H.c.}\right] \right\} \notag\\
+&\frac{1}{2}\sum_{i \neq j}V_{ij}\left[ \hat J^{(i)}_{a,z} \hat J^{(j)}_{a,z}\!-\!\frac{1}{4}\left(\hat J^{(i)}_{a,+}  \hat J^{(j)}_{a,-}\!+\!\mathrm{H.c.}\right)\right]\,.
\label{eq:Hfull}
\end{align}
The first line describes linear and nonlinear single-particle terms. 
As previously mentioned, the $\kappa = 1$ term can also be obtained by MW engineering ($\lambda_1 = \Omega_a$), see Eq.~\eqref{eq:MW}.
The second line describes long-range dipolar interactions under the assumption that the atoms are placed in a plane perpendicular to the static electric and magnetic fields ($\theta_{ij} = \pi/2$), thus realizing a `large-spin' XXZ model. 

In Sec.~\ref{sec:SGmodel}, we will discuss two examples of specific many-body models obtained from the Hamiltonian~\eqref{eq:Hfull} that provide a direct application of the platform developed in this work. 
State preparation and readout for the $\mathcal H^a$ manifold require specific schemes involving the states in $\mathcal H^v$, which we discuss in detail in the next Subsection.

\subsection{Vertical manifold \texorpdfstring{$\mathcal H^v$}{Hv}}
\label{subsec:vertical}

\begin{figure}[!t]
\center
\includegraphics[width=1\columnwidth]{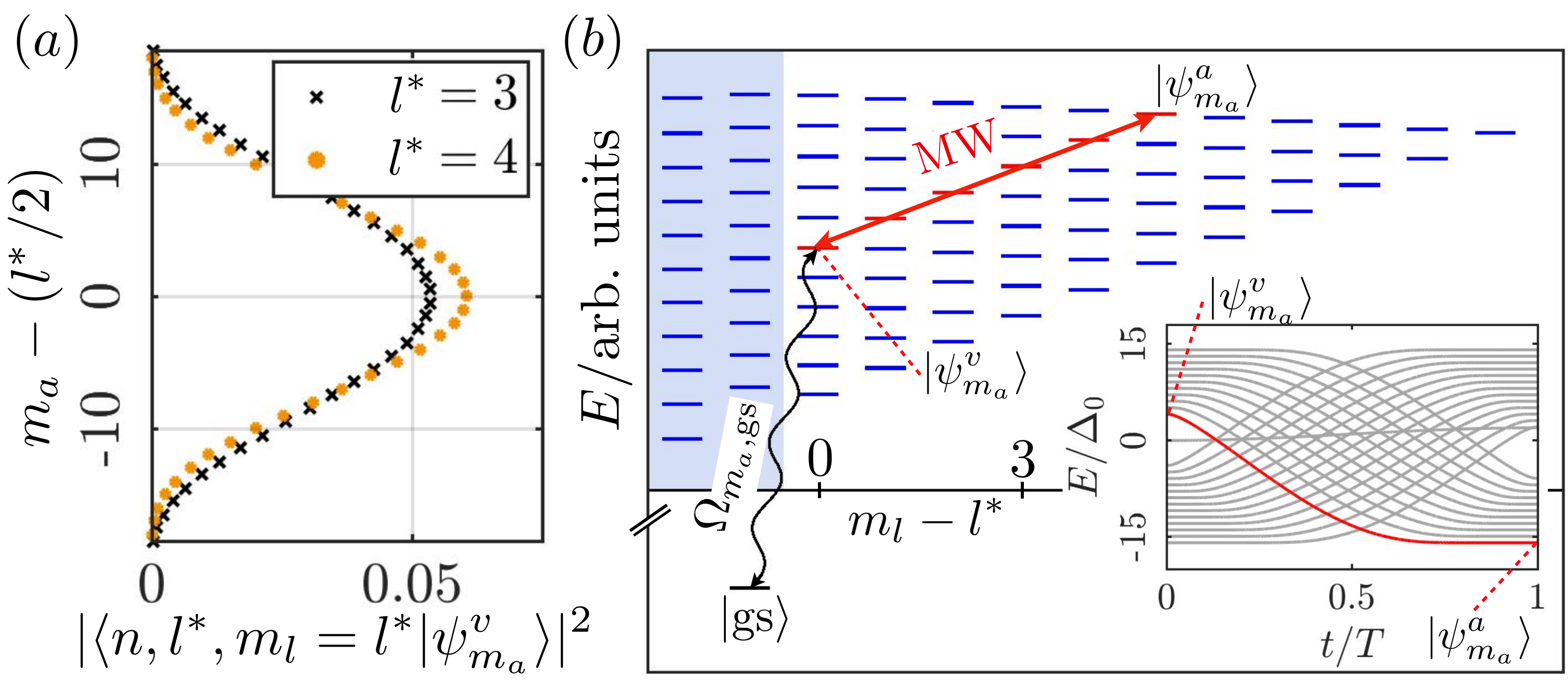}
\caption{State preparation and readout: (a) Low orbital angular momentum state admixture $|\langle n,l=l^*,l^*\ket{\psi^v_{m_a}}|^2$ \cite{Biedenharn1984} for the subset of states $\mathcal{H}^v$, where $\ket{n,l,m_l}$ are the eigenstates of the Hydrogen Hamiltonian $\hat H_H$, with $l$ being the orbital angular momentum quantum number. 
(b) Level structure of non-hydrogenic atoms in the presence of parallel electric and magnetic fields, with $\omega_Z=3\omega_S/4$. 
The blue shaded area displays the region modified by quantum defects. 
A ground (or intermediate-state) $\ket{\mr{gs}}$, with orbital angular momentum components $l^*-1$, is coupled with circularly polarized laser light to states in the sub-manifold $\mathcal{H}^v$.
The red arrow illustrates the multi MW photon coupling of the adiabatic rapid passage for a single value of $m_a$.
The inset shows the instantaneous eigenenergies of the adiabatic rapid passage path in the rotating frame defined above Eq.~\eqref{eq:MW} for a single $m_a = 0$ and for parameters $n=31$, $F = F_\mr{IT}/2$, $\omega_Z = \omega_S/2$, $\Delta_0 = \Omega_0>0$.}
\label{fig:RAP}
\end{figure}

The vertical manifold $\mathcal H^v$ defined in Eq.~\eqref{eq:vmanif} and highlighted in green in Fig.~\ref{fig:Fig1}c plays a fundamental role in our system, for multiple reasons.
First, this manifold can be accessed from atomic ground states using a single or multiple laser transitions (depending on the atomic species),
thus enabling to coherently (de)populate $\mathcal H^v$.
Second, a mapping scheme that we demonstrate below allows to transfer arbitrary states from $\mathcal H^v$ to $\mathcal H^a$ and vice versa with high fidelity.
As a consequence, the manifold $\mathcal H^v$ provides the capabilities to perform state preparation and readout on $\mathcal H^a$, which make the Hamiltonian in Eq.~\eqref{eq:Hfull} a particularly accessible many-body model.

The presence of low orbital angular momentum components in $\ket{\psi^v_{m_a}}$, see Fig.~\ref{fig:RAP}a, offers a unique opportunity for laser coupling the atomic ground state to the Rydberg manifold $\mathcal{H}^v$.
In particular, for Lanthanides like Erbium the valence electrons have $f$-orbital character in their electronic ground state due to a submerged shell structure \cite{Trautmann2021}. 
The $\mathcal{H}^v$ manifold of Erbium in turn contains an admixture of Rydberg $g$-states, which allows direct laser coupling from the ground state $|\mr{gs}\rangle$ to specific states in $\mathcal{H}^v$, selected by the laser frequency, see Fig.~\ref{fig:RAP}b. 
For multiple laser frequencies, this coupling is described in the rotating frame by the Hamiltonian
\begin{align}
    \label{eq:holo}
    \hat H^{vg} = \sum_{m_a} \left( \frac{\Omega_{m_a, \mr{gs}}}{2}\, |\psi_{m_a}^v\rangle \langle \mr{gs}| + \mr{H.c.} \right) \,,
\end{align}
where $\Omega_{m_a, \mr{gs}}$ are the individual Rabi frequencies that satisfy $|\Omega_{m_a, \mr{gs}}|\ll |\omega_S|$.
This laser adressability can also be used to perform projective readout within the $\mathcal{H}^v$ manifold employing quantum gas microscope techniques \cite{Bakr2009, Sherson2010}.
Note that the connectivity offered by the individual $\Omega_{m_a, \mr{gs}}$ in Eq.~\eqref{eq:holo} provides a convenient setting to perform holonomic quantum computing operations within the $\mathcal{H}^v$ manifold \cite{Zanardi2012}.

We now proceed to discuss a scheme realizing the state transfer
\begin{align}
    \hat U^{va}:\,\, |\psi_{m_a}^v\rangle \longleftrightarrow |\psi_{m_a}^a \rangle \quad \forall m_a \,,
\end{align}
via a adiabatic rapid passage that generalizes the one originally developed to prepare circular Rydberg levels~\cite{Hulet1983}. 
We start by noting that a state $\ket{\psi^v_{m_a}}$ is connected to the final state $\ket{\psi^a_{m_a}}$ for any $m_a$ by a multi-photon MW transition (see the highlighted red arrow and levels in Fig.~\ref{fig:RAP}b), while transitions into the defect region are energetically suppressed.
The required time-dependent drive is based on the MW Hamiltonian $\hat H_\mr{MW}(t)$ (Eq.~\eqref{eq:MW}) with $\Omega_a \!=\! 0$, such that only couplings among the highlighted states in Fig.~\ref{fig:RAP}b are present.
During the driving sweep, the detuning $\Delta_b(t)$ has to change sign and the Rabi coupling $\Omega_b(t)$ has to be turned on and off again in order to adiabatically ($|\Omega_b(t)|\ll |\omega_{a(b)}|$) connect the different states $\ket{\psi^v_{m_a}}$ and $\ket{\psi^a_{m_a}}$.
Note that in general the individual states acquire a state-dependent dynamical phase that can be canceled by an appropriate transformation in $\mathcal H^v$.

In the following, we numerically demonstrate the adiabatic rapid passage $\hat U^{va}$ for a selected state ($m_a = 0$) and a particular driving protocol $\Delta_b(t) = \Delta_0\cos(\pi\,t/T)$ and $\Omega_b(t) = \Omega_0\sin(\pi\,t/T)$.
In the inset of Fig.~\ref{fig:RAP}b, we show the instantaneous energy levels in the rotating frame defined above Eq.~\eqref{eq:MW} as a function of time.
For $t=0$, several energy levels are present. 
The ones with positive energy ($E>0$) are equally spaced and correspond to the levels highlighted in red in Fig.~\ref{fig:RAP}b.
Levels with different $m_a$ are not shown, as they are not coupled by the MW field.
As the adiabatic sweep progresses, the initial state $\ket{\psi^v_{m_a}}$ changes into the target final state $\ket{\psi^a_{m_a}}$, as indicated by the red line in the inset.
However, one can notice that several level crossings take place during the time evolution. 
These occur with states that have negative energy $E<0$ at $t=0$, corresponding to states with $m_l < -l^*$ and the same value of $m_a$, i.e. located on the left side of the rhombus level structure.
Additionally, level crossings with states from the defect modified region ($|m_l|<l^*$) can also be present.
All these crossings are in fact avoided crossings with a very narrow gap that is determined by multiple off-resonant transitions through the defect region. 
Thus, they can be diabatically crossed via Landau-Zener tunneling without hindering the sweep protocol.

Due to the highly regular Rydberg level spacing, we remark that the passage connects all the states in $\mathcal{H}^v$ to their respective counterpart in $\mathcal{H}^a$. 
For $\Delta_0$ and $\Omega_0$ on the order of $2\pi\times 10\,\mathrm{MHz}$ and $T$ on the order of a few $\mu\mr{s}$, the passage reaches numerically calculated fidelities above $99\%$ for all the states.

In summary, the hydrogen-like Rydberg manifold discussed in this paper can be accessed by laser exciting atoms to Rydberg states with small orbital angular momentum components and a subsequent transfer to states with large orbital angular momentum components.
This allows to explore the physics offered by the engineered many-body models introduced in Sec.~\ref{subsec:triangular} and \ref{subsec:edge}. 
In the following Sections we show how these models can be used to simulate Quantum Field Theories and how controlled entanglement between pairs of atoms can be generated.

%%%%%%%%%%%%%%%   sine-Gordon model %%%%%%%%%%%%%%%%
\section{Application: Quantum simulation}
\label{sec:SGmodel}
The central feature of the simulator discussed in this work is the high-dimensional and regularly structured Rydberg manifold leading to a many-body problem described by Heisenberg models with large spins (or angular momenta). 
A promising application of these models is the simulation of Quantum Field Theories (QFTs) \cite{jordan2012quantum}, which we discuss in this Section.
For instance, a scalar quantum field is represented by pairs of conjugate variables on each spatial point, which span an infinite-dimensional space. 
These can be conveniently represented by the large spins of the Rydberg $n$-manifold, a procedure that naturally requires a truncation set by the finite (large) spin length.

Our construction provides a bottom-up approach to simulating QFTs, differently from other systems, e.g. ultracold atomic gases \cite{Schweigler2017}, where QFTs appear as a low-energy description.
The Rydberg platform offers the opportunity to avoid unwanted thermal effects inherent to atomic gases, thus allowing us to access the zero temperature limit of these theories and to freely choose different geometries or dimensionalities.
Moreover, the programmability and control of the platform allows for the preparation of initial states of interest, e.g. product states or other states obtained through adiabatic sweeps, the readout of the corresponding out-of-equilibrium quantum dynamics via quench protocols and the exploration of phase diagrams from the weakly- to the strongly-interacting regime. 
These are notoriously hard computational problems that the Rydberg simulator can tackle.

In this Section, we illustrate these possibilities with two case studies, namely the massless and massive sine-Gordon model.
The first case is a paradigmatic and ubiquitous integrable scalar QFT that plays an important role in many areas, ranging from high-energy physics \cite{coleman1975quantum} to condensed matter \cite{Giamarchi2003}.
The second case is a scalar QFT that is dual to 1+1D Quantum Electrodynamics, namely the massive Schwinger \cite{coleman1976more}. 
The quantum simulation of such a theory requires a continuous variable representation of the scalar fields, which are naturally embedded in the Rydberg manifold, as we show below. 

In Subsec.~\ref{subsec:CV}, we first discuss how the platform's unique properties, namely the availability of spins with large spin length $J \!\gg\! 1$, can be used to faithfully approximate the physics of continuous variables. 
In the following Subsec.~\ref{subsec:SG}, we show that the naturally occurring interactions enable us to realize a lattice regularization of the sine-Gordon (SG) model.
Finally in Subsec.~\ref{subsec:QED}, we turn our attention to gauge field theories, more precisely quantum electrodynamics in one spatial dimension, the so-called massive Schwinger model. 
We discuss how to realize a dual lattice formulation of this model using the Rydberg architecture and demonstrate how to probe nontrivial gauge theory dynamics, relevant for investigating processes like pair production and string breaking.

\subsection{Conjugated continuous variables
\label{subsec:CV}}
Standard quantum-mechanical position $\hat{\varphi}$ and momentum $\hat{\pi}$ operators obey the canonical commutation relation $\left[\hat{\varphi}, \hat{\pi}\right] = i$.
For quantum mechanics on a unit circle, it will be more convenient to consider phase operators $e^{i\hat{\varphi}}$ satisfying
\begin{align}\label{eq:comm_rel_phase}
    \left[\hat{\pi},e^{i\hat{\varphi}} \right] = e^{i\hat{\varphi}} \;,
\end{align}
where the momentum operator $\hat \pi$ takes discrete eigenvalues $m_\pi \in \mathbb{Z}$ on its eigenstates $|m_\pi\rangle$. 

The commutation relation ~\eqref{eq:comm_rel_phase} shows that $e^{i\hat{\varphi}}$, similar to a spin operator $\hat{J}_+$, acts as a raising operator on $|m_\pi\rangle$. Taking into account the normalization of the operators, this suggests to identify
\begin{align}\label{eq:CV_spin_identification}
    \hat{\pi} \leftrightarrow \hat{J}_z \;, && e^{\pm i\hat{\varphi}} \leftrightarrow \frac{1}{\sqrt{J(J+1)}}\hat{J}_\pm \;.
\end{align}
The $(2J+1)$-dimensional Hilbert space spanned by eigenstates $|m\rangle$ of $\hat{J}_z$ with $m= -J, \dots J$ thus ``regularizes'' the infinite-dimensional Hilbert space of the continuous variables by cutting off the spectrum of $\hat{\pi}$ at a maximum value $|m_\pi| \le J$~\cite{haldane1983continuum}. 
The identification becomes exact for states around $m=0$ in the limit $J\gg 1$.

We now numerically demonstrate how the above identification converges to a continuous variable representation in the large spin limit $J\gg 1$. 
Here, we focus on a single Rydberg atom in the subset of states $\mathcal{H}^a$ described by the Hamiltonian in Eq.~\eqref{eq:Hfull}.
Choosing $\Delta_a = 0$ and defining $\Omega^\prime = -2\Omega_a \sqrt{J(J+1)}$ and $\lambda' = -2 \lambda_\kappa e^{i \theta} [J(J+1)]^{\kappa/2}$ for a fixed $\kappa>1$, $\lambda'>0$ and a phase $\theta$, the single Rydberg atom Hamiltonian in the continuous variable limit becomes
\begin{equation} \label{eq:singleatom_cont}
    \hat H_{J\rightarrow\infty} = \chi \, \hat \pi^2 - \Omega^\prime \cos \hat \varphi - \lambda^\prime \cos(\kappa\hat\varphi+\theta) \;.
\end{equation}
 
\begin{figure}
    \centering
    \includegraphics[width=0.99\columnwidth]{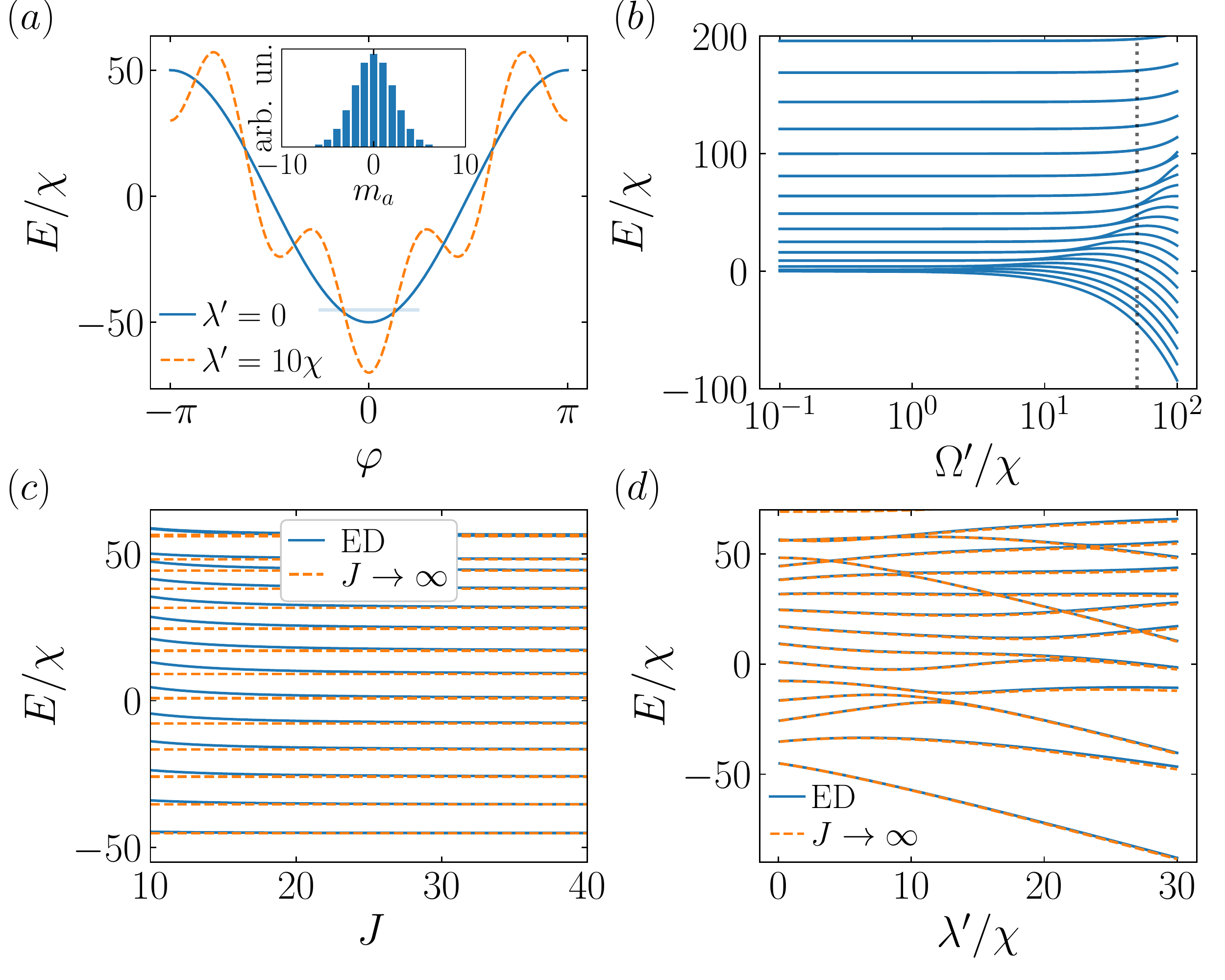}
    \caption{
    Large spin limit for a single Rydberg atom. (a) Potential landscape of the continuum theory with Ponderomotive coupling $\lambda'=0$ (solid line) and $\lambda'=10\chi$ (dashed line). 
    The horizontal line marks the ground state energy for $\lambda'=0$ and $\Omega'=50\chi$ in the limit $J\rightarrow\infty$.  In the inset, the corresponding wavefunction is shown in momentum space. 
    (b) Continuum theory energy spectrum for $\lambda'=0$. The vertical dashed line indicates the value $\Omega'=50\chi$ used in the rest of the figure. 
    (c) Spectrum of the single-particle Rydberg Hamiltonian (solid lines) compared with the exact continuum theory spectrum (dashed lines) as $J$ increases.
    (d) Spectrum comparison as in $(c)$ for $J=50$ as a function of the Ponderomotive coupling $\lambda'/\chi$ for $\kappa = 4$.}
    \label{fig:largeJ}
\end{figure}

The Hamiltonian in Eq.~\eqref{eq:singleatom_cont} describes a particle of mass $(2\chi)^{-1}$ moving on a ring and subject to a periodic potential, as depicted in Fig.~\ref{fig:largeJ}a.
The spectrum of the continuum theory \eqref{eq:singleatom_cont} can be numerically computed by exploiting the periodicity of $\hat \varphi$ and solving the eigenvalue problem in Fourier space, thus yielding the result shown in Fig.~\ref{fig:largeJ}b for $\lambda^\prime=0$.
Depending on the ratio $\Omega^\prime/\chi$, the low-energy physics of this model ranges from the free particle regime ($\Omega^\prime \ll \chi$) to the harmonic oscillator regime ($\Omega^\prime \gg \chi$). 

We now consider an intermediate regime, where the nonlinear character of $\cos\hat\varphi$ is evident, and calculate the spectrum of the spin model for increasing values of the angular momentum $J$, as shown in Fig.~\ref{fig:largeJ}c. 
For the chosen parameters and the selected energy range, the truncation accurately reproduces both bound and a few unbound states of the model with continuous variables, Eq.~\eqref{eq:singleatom_cont}, when $J \gtrsim 30$. 
In Fig.~\ref{fig:largeJ}d, we further test the effect of $\lambda^\prime$ on the spectrum for a fixed value of the spin magnitude $J=50$, and find that the spin model also captures the target model including the higher-harmonic potential, controlled by $\lambda^\prime$.

We have thus verified that for experimentally relevant spin lengths $J$ (or principal quantum numbers $n=2J+1$) the large spin captures the physics of a continuous variables theory in a single Rydberg atom.
This capability constitutes the basis to engineer a QFT simulator using an array of interacting Rydberg atoms, which we illustrate in the following Subsection for the SG model.

\subsection{Sine-Gordon model
\label{subsec:SG}}

\subsubsection{Overview of the model and Rydberg implementation}

The SG model is a paradigmatic QFT that is described by the Hamiltonian~\cite{coleman1975quantum,abdalla1991non}
\begin{align}\label{eq:cont_SG}
    \hat{H}_\mr{SG} \!=\!  \int \mr{d} x \, \bigg\{\frac{1}{2}[\hat\pi(x)]^2 \!+\! \frac{1}{2}[\partial_x\hat\varphi(x)]^2
    -\frac{M_0^2}{\beta^2}\cos[\beta \hat\varphi(x)]\bigg\} \;,
\end{align}
where the fields obey $[\hat \varphi(x),\,\hat\pi(x')]=i\delta(x-x')$, $M_0$ has units of energy, distances are measured in units of inverse energy and $\beta$ is a dimensionless parameter.
Let us briefly summarize some key features of the SG model (see, e.g., \cite{sabio2010sudden}) and references therein).
At $\beta^2 = 8\pi$, the theory exhibits a Berezinskii-Kosterlitz-Thouless (BKT) transition, separating a gapless phase at large $\beta$ from a gapped phase at small $\beta$.
In the gapped phase ($0 \!<\! \beta^2 \!<\! 8\pi$), the low-energy degrees of freedoms are fermions interacting attractively (repulsively) for \mbox{$\beta^2 < 4\pi$} (\mbox{$\beta^2 > 4\pi$}), which can be identified with quantized solitons of mass $M \overset{\beta\rightarrow 0}{\longrightarrow} 8 M_0/\beta$.
In the attractive regime, the solitons form bound states, so-called breathers. The lightest one of them has a mass $m_1\overset{\beta\rightarrow 0}{\longrightarrow} M_0$
(see App.~\ref{Appendix:SGExact} for general expressions of the masses).

For the implementation of the SG model we consider a one-dimensional array of tweezer-trapped Rydberg atoms excited to the the manifold $\mathcal H^a$. 
The system is therefore described by the many-body Hamiltonian~\eqref{eq:Hfull}, which in the large $J$ limit allows for a continuous variable description through Eq.~\eqref{eq:CV_spin_identification} that reads 
\begin{align}
    \label{eq:SGlat}
    \hat{H}_{\text{SG}}^{(\text{lat})} \!=\! &\sum_i \left[\chi \hat{\pi}_i^2 -  \lambda' \cos(\kappa\hat{\varphi}_i)  - \sum_{j> i}\frac{V_{ij}'}{2}\cos \left(\hat{\varphi}_{j}\!-\!\hat\varphi_i\right) \right] \,,
\end{align}
where we rescaled the interaction \mbox{$V_{ij}' = V_{ij} J(J+1)$}, while keeping $\Delta_a=0$, and $\lambda' = -2\lambda_\kappa [J(J+1)]^{\kappa/2}$ for a single value of $\kappa$, which corresponds to either a MW ($\kappa=1$) or ponderomotive ($\kappa \geq 1$) coupling, respectively.
Notice that in the large $J$ limit, the Ising terms have been neglected, which we benchmark in the next Subsection.

In order to recover the continuum QFT in Eq.~\eqref{eq:cont_SG}, we take only nearest-neighbor terms $V'_{i,i+1} \equiv V'_\text{nn}$ and rescale the continuous variables and Hamiltonian appropriately (see App.~\ref{Appendix:SGimplementation}).
The correspondence is achieved by the identification of parameters as follows
\begin{align}
   \frac{V'_\text{nn}}{\chi} = \frac{4\kappa^4}{\beta^4}\;, &&   \frac{\lambda'}{\chi} = \frac{2\kappa^2(\ell M_0)^2}{\beta^4}  \;,
\end{align}
where we introduced a short-distance scale $\ell$ that sets a UV-cutoff $\Lambda \propto 1/\ell$ for the lattice regularization, \mbox{$\ell M_0 =\kappa \sqrt{2\lambda'/V'_\text{nn}} \rightarrow 0$}.
With these identifications, all the regimes of the theory,
\mbox{$0 < \beta^2 =2\kappa^2 \sqrt{\chi/V'_\text{nn}} < \infty$}, are in principle accessible.
In particular, in App.~\ref{Appendix:SGimplementation} we present realistic experimental parameter estimates, which indicate that values $\beta^2 \gtrsim 8\pi$ can be reached.
This suggests that this platform can be employed to investigate the critical properties of the model across the quantum phase transition.

Before turning to a benchmark analysis of our SG model implementation, let us briefly point out some existing proposals and experimental realizations.
The SG model can be realized with ultracold atoms, for example, by using binary mixtures ~\cite{Son2002, Recati2022} or tunnel-coupled superfluids \cite{Gritsev2007}, and the theory has been tested in the semiclassical regime by measuring high-order correlation functions~\cite{Schweigler2017}. 
Recently, it has been proposed to realize a lattice regularization of the quantum SG model using superconducting circuits~\cite{Roy2021}, which is most closely related to our approach with large spins discussed here.

\begin{figure}
    \centering
    \includegraphics[width=0.99\columnwidth]{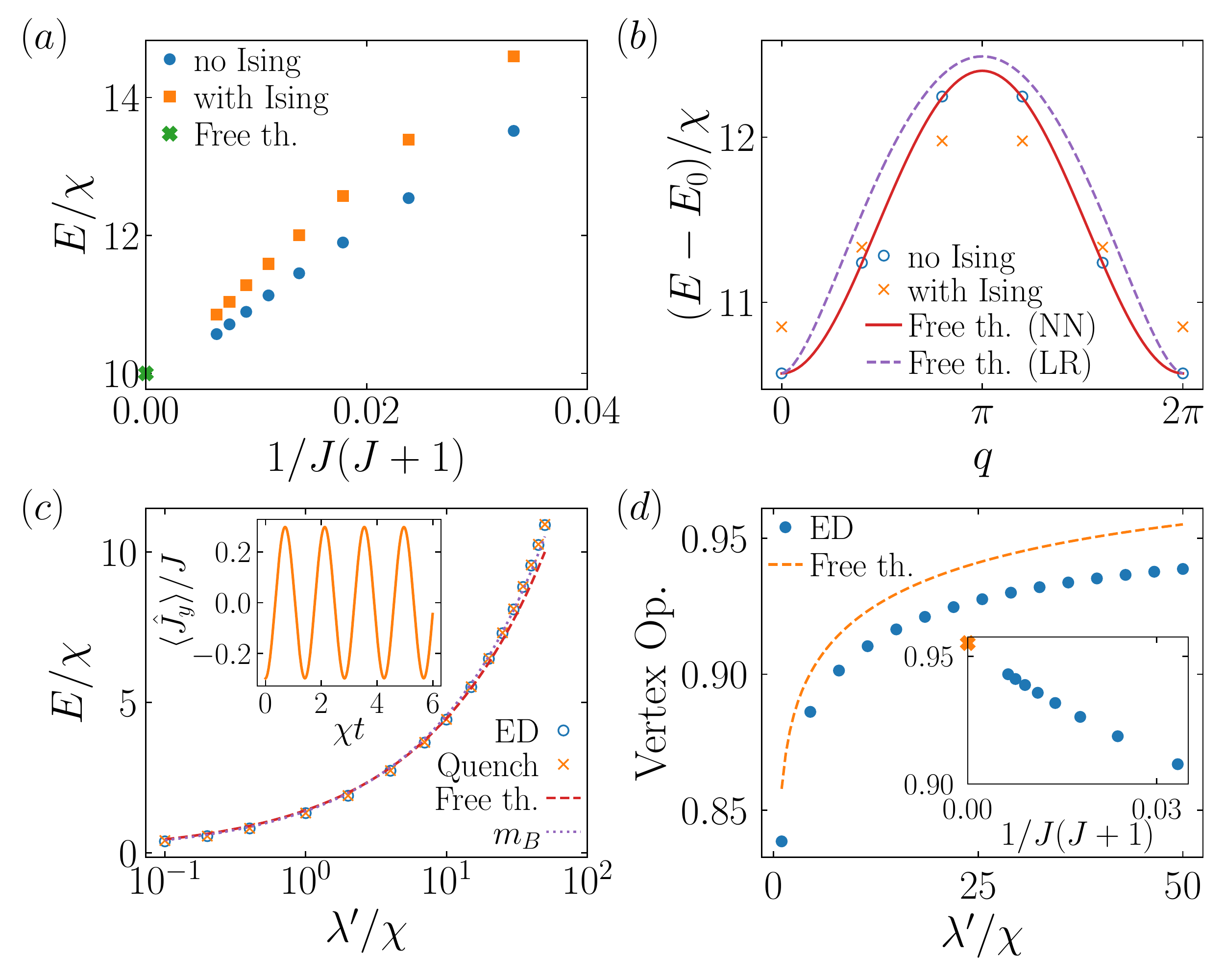}
    \caption{Sine-Gordon benchmark. (a) Spin-size scaling of the energy gap up to $J=12$ obtained from the spin Hamiltonian including the Ising terms (full circles) and without them (full squares) for $N=5$ atoms, $\kappa = 1$  $\lambda'=50\chi$ and $V_{\text{dd}}' = 10 \chi$ obtained by exact diagonalization with periodic boundary conditions (PBC). The prediction of the free theory is shown for the $J\gg1$ limit. (b) Lowest branch of the excitation spectrum obtained by ED (markers) with the parameters chosen as in (a) for $J=10$. The free theory prediction is shown in the nearest-neighbors (NN) approximation (dashed line) and including the entire long-range (LR) tail of the interactions (dotted line). (c) Energy gap for a wide range of $\lambda'/\chi$ obtained via the quench protocol with $\alpha = \pi/50$ described in the text (cross markers) and compared with the ED gap (empty circles). The plot shows also the free theory prediction (dashed line) and the exact lowest breather mass $m_B=m_1$ (dotted line). In the inset, the oscillation dynamics of $\langle \hat J_y \rangle$ from which we extract the gap for $\lambda'=10\chi$. (d) Vertex operator expectation value $\langle e^{i\hat \varphi} \rangle \leftrightarrow  \langle \hat J_x \rangle /\sqrt{J(J+1)} $ obtained with ED (full circles) and within the free theory (dashed line). The discrepancy is mainly originating from the small spin length, as shown in the inset for $\lambda'=50\chi$. ED in panels (b)-(d) is performed for $J=10$ and the site indices are dropped as we considered a translation invariant system.
    }
    \label{fig:SG_illustration}
\end{figure}

\subsubsection{Numerical benchmark in the massive phase}
We now illustrate how the lattice sine-Gordon physics emerges from the large spin approximation. 
In particular, we test the influence of finite spin length, the presence of Ising terms and discuss the effect of long-range tails.
For this, we numerically analyze an array of $N=5$ Rydberg atoms assuming periodic boundary conditions \cite{Labuhn2016} and compare with analytical results for the low-energy excitation spectrum and ground state expectation values.
Further below, we also briefly discuss how to measure the many-body gap through a realistic quench protocol.

We perform our analysis deep in the gapped phase, setting $V'_\text{nn} = 10 \chi$ and $\kappa=1$, \emph{i.e.} $\beta^2 = \sqrt{0.4} \ll 8\pi$, and vary $\lambda'/\chi$, thereby effectively scanning different values of  $\ell M_0$.
In this parameter regime, we expect the model to be well approximated by a quadratic (free) theory obtained by replacing $\cos \hat \varphi_i \approx 1 \!-\! \hat \varphi_i^2/2$, and similarly for $ \cos \left(\hat{\varphi}_{i+1} - \hat\varphi_i\right)$. 
After diagonalizing the quadratic theory (see, e.g., \cite{Bruus2004} and App.~\ref{Appendix:FreeTheory}), we obtain the single-particle dispersion relation \mbox{$\omega_q = \sqrt{2\chi(\lambda^\prime+V^\prime_{\text{nn}}-V^\prime_{\text{nn}}\cos q)}$} with $q \in [0,2\pi]$, where we truncated the interactions to nearest neighbors. 
These low-energy massive excitations with dispersion $\omega_q$ correspond to coherent displacements of the phase variable from the equilibrium position $\langle \hat \varphi_i \rangle = 0$, which can be interpreted as gapped spin waves of the original spin model.

As illustrated in Fig.~\ref{fig:SG_illustration}a for finite $J$, the numerically calculated dispersion obtained by neglecting Ising terms and long-range contributions is in perfect agreement with the analytical formula for $\omega_q$.
We find that including the Ising terms has the effect of reducing the overall bandwidth of the dispersion relation. 
In Fig.~\ref{fig:SG_illustration}a, we also show the effect of the long-range dipolar tail, which is calculated semi-analytically in App.~\ref{Appendix:FreeTheory} by using Ewald's resummation \cite{Peter2015}.
For \mbox{$q\approx 0$}, we find \mbox{$\omega_q \approx \sqrt{2\chi[\lambda'- V_\text{dd}'(c_2^{} q^2 + c_2^\prime\, q^2 \log q)]}$}, with $c_2 \approx -0.739$ and $c_2'\approx 1/2$. 
The long-range dipolar interactions have therefore the effect of increasing the bandwidth of the excitation spectrum, which gains a non-analytical sub-leading contribution at small momenta.
Note that the energy gap \mbox{$\Delta E = \omega_0 = \sqrt{2\chi\lambda^\prime}$} is thus unaffected by the long-range tails.

We now  study the gap $\Delta E$ of the model in more detail. 
As previously stated, we expect the Ising terms' contribution to vanish in the $J\gg 1$ limit. 
This is confirmed in Fig.~\ref{fig:SG_illustration}b, where we show the dependence of the energy gap $\Delta E$ in the finite-spin model as a function of $J$ with $J=5, \dots, 12$, both with and without Ising terms.
Comparing against the predicted result $\omega_0=\sqrt{2\chi\lambda'}$ of the free theory, we find that both finite-spin results overestimate the value of the gap.
These systematic errors vanish smoothly as $J$ is increased, demonstrating that the finite spin model reproduces the expected gap in the large $J$ limit.
Given the regular $J$-dependence, we note that a finite spin-length scaling analysis can also be performed in an experimental realization by repeating appropriate measurements in different $n$ manifolds.
In principle, this allows to extrapolate the asymptotic value of the gap, thereby improving the prediction of the Rydberg quantum simulator.

In a quantum simulation experiment based on the Rydberg platform discussed in this work, the gap can be accessed by the following quench protocol.
First, we prepare the ground state of Eq.~\eqref{eq:SGlat} with \mbox{$\cos\hat \varphi_i \rightarrow \cos(\hat \varphi_i+\alpha)$}, which can be obtained in the spin model by taking, for example, a MW field in Eq.~\eqref{eq:MW} with complex Rabi frequency $\Omega_a \rightarrow \Omega_a e^{i\alpha}$.
For small $\alpha$, this corresponds to a small uniform displacement of the spin expectation value away from $\langle \hat J_y^{(i)} \rangle=0$.
Quenching to $\alpha = 0$, the energy gap can be extracted by measuring the oscillation frequency of $\langle \hat J_y^{(i)} \rangle (t)$.

The leading contribution of the induced dynamics within linear-response, i.e. for \mbox{$|\alpha| \!\ll\! \pi$}, is \mbox{$\langle \hat J_y^{(i)} \rangle \propto \langle \hat \varphi_i \rangle + O\left(\langle \hat \varphi_i^3 \rangle \right)$}, which therefore gives direct access to the coherent displacement of the phase variable. 
The determination of the many-body gap can be accurately determined for a wide range of $\lambda^\prime/\chi$ values, as shown in Fig.~\ref{fig:SG_illustration}c.
The initial state preparation can be performed using the scheme outlined in Sec.~\ref{subsec:vertical} for large $\lambda'$ followed by an adiabatic ramp to the final value of $\lambda'/V'_\mr{nn}$ and $\lambda'/\chi$, which is protected by the many-body gap.
The same scheme in Sec.~\ref{subsec:vertical} also allows to measure the value of $\langle \hat J_y^{(i)} \rangle$ after the time evolution.

Several other types of correlation functions are also measurable via the scheme detailed in Sec.~\ref{subsec:vertical}. 
We give another example in Fig.~\ref{fig:SG_illustration}d, where we compute the ground-state expectation value of the vertex operator via $\langle \hat J_+^{(i)} \rangle \propto \langle e^{i\hat\varphi_i}\rangle =\langle \cos\hat\varphi_i\rangle$ and compare it against the free theory prediction $\langle \cos\hat\varphi_i\rangle =\exp\left[ -\frac{\chi}{2N}\sum_q \frac{1}{\omega_q} \right]$. 
The plot displays a systematic difference between the numerical ED result and the free theory prediction. As shown in the inset for $\lambda' = 50\chi$, this is mainly originating from the finite spin length ($J=10$) used in the ED simulations. 
Note that the observables discussed so far can be measured by only using global MW control.
However, more general correlation functions as $\langle \hat J_+^{(i)} \hat J_-^{(j)} \rangle$, with $i\neq j$, require single site control, which can be achieved by using electric field spatial gradients for MW fields \cite{Daley2008} or local laser addressing, see Sec.~\ref{subsubsec:Pomderomotive}.

The main discussion of this Section has focused on benchmarking low-energy SG properties in the massive phase. 
However, the Rydberg simulator offers the unique opportunity to prepare more general high-fidelity pure states and study their non-equilibrium quantum dynamics, e.g. after quenching the model parameters.
An intriguing property of the SG model is the presence of topological excitations, or solitons, which are relevant to characterize the BKT transition.
A possible application is the simulation of such high-energy metastable states \cite{Fialko2015, Abel2021, Milsted2022}. This will provide a testbed to study questions related to false-vacuum decay \cite{Coleman1977}, which plays a central role in cosmological models as well as in first-order phase transitions \cite{Song2022}.

\subsection{Massive Schwinger model}
\label{subsec:QED}

\subsubsection{Overview of the model and Rydberg implementation}

We now come to our final quantum simulation application, namely Quantum Electrodynamics (QED) in one spatial dimension, also known as the massive Schwinger model. 
Here we make use of the fact that a non-integrable modification of the SG model, the so-called massive SG model, described by the Hamiltonian
\begin{align}
\label{eq:H_QED}
    \hat H_{\text{mSG}}= \int \mr{d}x \, \bigg\{\frac{1}{2} [\hat \pi(x)]^2 &+ \frac{1}{2} [\partial_x \hat\varphi(x)]^2 + \frac{M^2}{2} [\hat \varphi(x)]^2 \nonumber \\
    & - u \cos [\beta \hat \varphi(x) + \theta]\bigg\} \;,
\end{align}
provides a dual low-energy effective description of the massive Schwinger model~\cite{coleman1976more,abdalla1991non,jentsch2022physical}, described by
\begin{align}
    \hat H_\text{QED} = \int \mr{d}x \, \bigg\{ \frac{1}{2} &\left[\hat E(x) + \frac{e\theta}{\pi}\right]^2 \\ 
    &+ \hat {{\psi}}^\dagger(x) \gamma^0 \left[-i \gamma^1 \hat D_x + m \right] \hat \psi(x) \bigg\} \;. \nonumber
\end{align}
Here, $\gamma^{0(1)}$ are the gamma matrices in two space-time dimensions, we abbreviated $\hat D_x = \partial_x - ie\hat A(x)$, the fermionic spinors obey $\left\{\hat\psi_j(x), \hat \psi^\dagger_{j'} (x')\right\} = \delta_{jj'} \delta(x-x')$ and the gauge fields fulfill $\left[\hat A(x), \hat E(x')\right] = i \delta(x-x')$.
The duality requires to fix the mass $M$, the coupling $u$ and the parameter $\beta$ as
\begin{align}
    \beta = 2\sqrt{\pi} \;, && M^2 = \frac{e^2}{\pi} \;, && u = \frac{\exp({\gamma})}{2\pi} \Lambda m \;,
\end{align}
where $e$ denotes the electric charge, $m$ is the free fermion mass, and $\gamma \approx 0.577$ is the Euler-Mascheroni constant. The above duality is valid in the continuum limit for a UV cutoff (imposed on the Hamiltonian \eqref{eq:H_QED}) \mbox{$\Lambda \gg e, m$}. 
In this limit, one obtains a QFT described by two dimensionless parameters, the coupling $0 \!\le\! e/m \!\le\! \infty$ and the topological angle \mbox{$\theta \in \left[-\pi,\pi\right]$}.

The Schwinger model is a well-studied toy model that shares several qualitative properties with more complicated gauge theories such as quantum chromodynamics, for example confinement.
Let us briefly discuss some key features of the model~\cite{coleman1976more}, starting with the case of $\theta = 0$.
For $e/m = 0$, the gauge fields trivially decouple and we obtain a theory of non-interacting massive fermions. 
For finite $e/m>0$, the gauge fields mediate a density-density interaction $\propto \hat \rho (x) \hat \rho (x')$ among the charge densities \mbox{$\hat \rho = e \hat \psi^\dagger \hat \psi$} with a Coulomb potential growing with distance as $\propto |x-x'|$.
Fermions of opposite charge thus interact attractively and are confined into bosonic bound states. 
In the dual theory these are directly described by the field $\hat \varphi$, which can be identified with the electric field $\hat E / e$. 
In the limit $e/m=\infty$, namely $u/M^2 = 0$, these bosons propagate freely with a mass given by $M$. 

\begin{figure}
    \centering
    \includegraphics[width=\columnwidth]{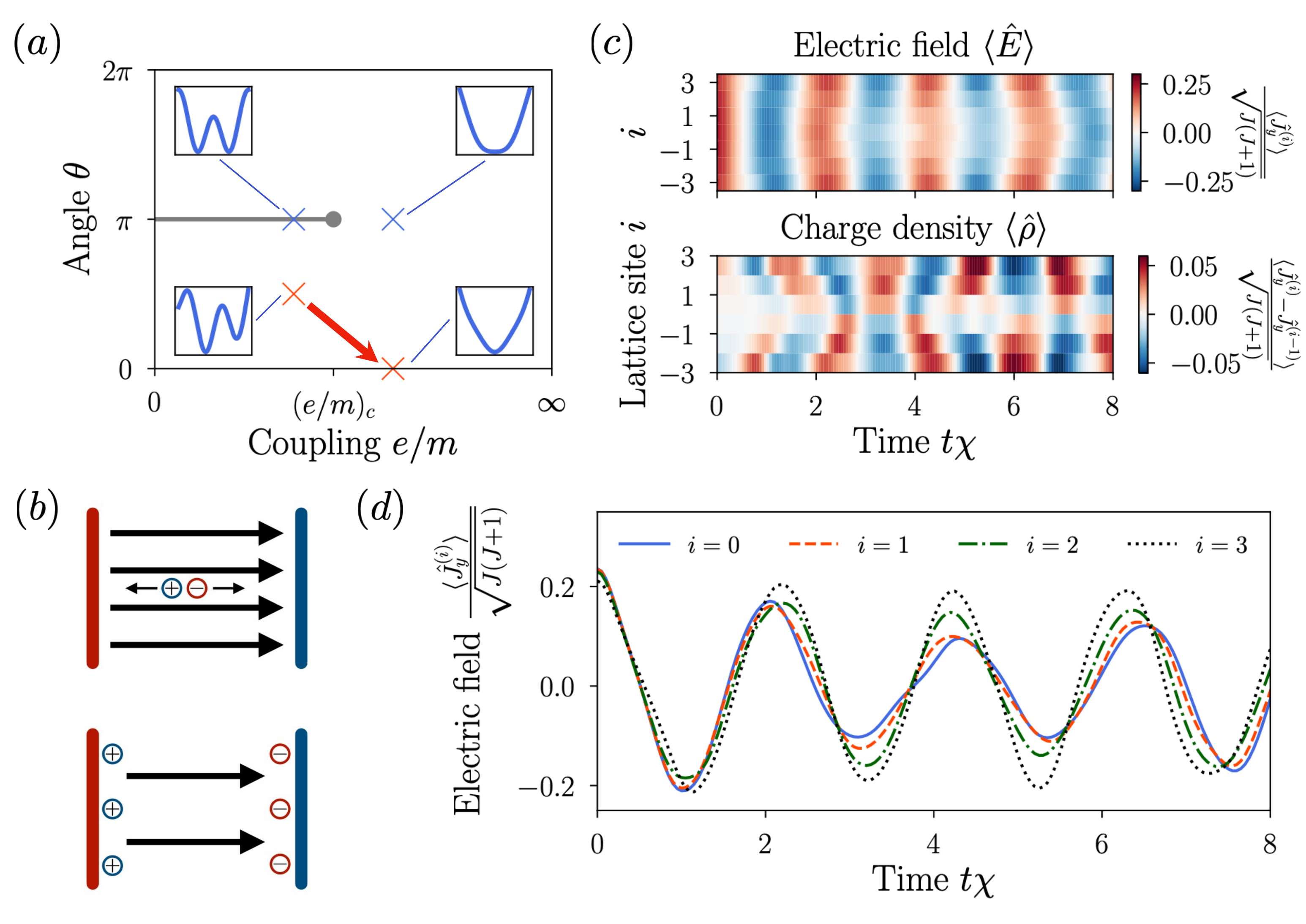}
    \caption{Schwinger model illustration. $(a)$ Phase diagram of the massive Schwinger model as a function of the angle $\theta \in [0,2\pi]$ and the coupling $e/m \in [0,\infty]$. The grey line at $\theta = \pi$ indicates the first-order phase transition, which terminates in a second-order critical end-point at $(e/m)_c$. The insets illustrate the shape of the interaction potential $V(\varphi) = \frac{M^2}{2} \varphi^2 - u \cos [\beta \varphi + \theta]$ at characteristic points indicated by the crosses and the red arrow marks the type of quench considered in panels $(c)$ and $(d)$. $(b)$ Sketch of capacitor plates with corresponding electric field that leads to the creation of charged particle pairs which in turns reduces the background electric field. $(c)$ Time evolution of the (lattice) electric field and charge density as measured by the expectation values $\langle \hat E_i \rangle \propto \langle \hat J^{(i)}_y\rangle$ and $\langle \hat \rho_i \rangle \propto \langle \hat J^{(i)}_y-\hat J^{(i-1)}_y\rangle$, respectively. $(d)$ Comparison of the electric field dynamics for the different sites. The electric field $\langle \hat E_i \rangle$ for the central lattice site $i=0$ exhibits a stronger damping than the outer sites $i=1,2,3$. In $(c)$ and $(d)$, the parameters of the simulation are given by $V'_\text{nn}/\chi = 4^4/(4\pi^2)$, $\Omega'/V'_\text{nn} = 0.5$, and $\lambda'/\chi = 2 \rightarrow 0.5$ is quenched simulatenously with $\theta = \pi/2 \rightarrow 0$. }
    \label{fig:QED_illustration}
\end{figure}

A finite value of $\theta$ corresponds to a background electric field, which is identical to the field created by two static background charges $\pm e \theta/ \pi$ via the Gauss law constraint $\partial_x \hat E = \hat \rho$. 
The physics is periodic in $\theta$ with period $2\pi$ (which is manifest in $\hat H_{\mr{mSG}}$) because any background charge $Q = \pm ne$ with integer $n$ is screened by the production of particle pairs induced by the corresponding background electric field, $\langle \hat E \rangle = Q$. 
For the special case of $\theta = \pm \pi$ (without additional external charges), the ground state of the model undergoes a second-order quantum phase transition at $(m/e)_c \approx 0.3335(2)$~\cite{byrnes2002density}, with order parameter $\langle \hat E \rangle$ (or $\langle \hat \varphi \rangle$), which falls in the Ising universality class and corresponds to the spontaneous breaking of the $\hat E \rightarrow - \hat E$ (or $\hat \varphi \rightarrow - \hat \varphi$) $\mathbb Z_2$ symmetry. 
For fixed $m/e > (m/e)_c$, the system undergoes a first-order phase transition at $\theta=\pm\pi$, see Fig.~\ref{fig:QED_illustration}a for a sketch of the phase diagram.
Further details can be found in Refs.~\cite{coleman1976more,jentsch2022physical}.

We obtain a lattice version of  Eq.~\eqref{eq:H_QED} by repeating a similar discretization as described in the previous Section for the SG model. 
This requires to include the MW field and the ponderomotive couplings simultaneously.
In particular, starting from the many-body Hamiltonian in Eq.~\eqref{eq:Hfull}, in the large $J$ limit the continuous variable description reads
\begin{align}
\label{eq:QED_J}
    \hat{H}_{\text{mSG}}^{(\mr{lat})} = 
    \sum_i \Big[\chi \hat{\pi}_i^2 &-  \lambda'_\kappa \cos(\kappa\hat{\varphi}_i+\theta) -\Omega'\cos(\hat{\varphi}_i)\notag\\
    &-\sum_{j> i}\frac{V_{ij}'}{2}\cos \left(\hat{\varphi}_{j}-\hat\varphi_i\right) \Big] \;,
\end{align}
where we use the identification \mbox{$V_{ij}' = V_{ij} J(J+1)$}, $\Omega' = -\Omega_a [2\sqrt{J(J+1)}]$ and $\lambda'_\kappa = -2\lambda_\kappa [J(J+1)]^{\kappa/2}$, and neglect the Ising interactions. 
Our target model is again approached in the continuum limit with the identifications 
\begin{align}
    \frac{V'_\mr{nn}}{\chi} = \frac{\kappa^4}{(2\pi)^2} \,, &&
    \frac{\lambda'_\kappa}{\chi} =
    \frac{\kappa^2\ell^2\exp(\gamma)\Lambda m}{(2\pi)^2}\,, && 
    \frac{\Omega'}{\chi} = \frac{\kappa^4e^2 \ell^2}{(2\pi)^3}\,,
\end{align}
and keeping only nearest-neighbor terms $V'_\mr{nn}\equiv V'_{i,i+1}$. 
As discussed in App.~\ref{Appendix:mSGimplementation}, the massive SG model in Eq.~\eqref{eq:H_QED} is recovered under the conditions 
\begin{align}\label{eq:sep_scales}
\chi,\, \lambda_\kappa' \ll \Omega' \ll V_\mr{nn}' \;,  \quad \lambda' \ll \sqrt{\chi V_\mr{nn}'}\,.
\end{align}

We emphasize that a suitable choice of parameters in principle allows us to probe the whole non-trivial range of the theory with $0 < e/m < \infty$.
A specific realistic example of parameters is provided in App.~\ref{Appendix:mSGimplementation} yielding $e/m \approx 4.5$.

In recent years, there has been a growing interest in quantum simulating gauge theories. Experimentally, real-time dynamics of the massive Schwinger model has been realized digitally on trapped-ion~\cite{martinez2016real} and super-conducting qubit~\cite{klco2018quantum} quantum computers. Analog simulations of directly related models have been performed with Rydberg arrays~\cite{bernien2017probing}, cold mixtures~\cite{mil2020scalable} and Bose-Hubbard systems~\cite{yang2020observation}. Our proposal differs from all of these existing realizations since we directly target the dual scalar field theory, which can be naturally implemented with Rydberg atoms in the $\mathcal H^a$ manifold.

\subsubsection{Numerical illustration: capacitor discharge}

In the remainder of this Section, we illustrate how our platform allows us to probe paradigmatic gauge-theory phenomena. 
Let us consider a system of size $D$ with a given background electric field determined by $\theta\neq 0$.
This effectively forms a charged capacitor with plates separated by distance $D$.
In general, the ground state of this system for sufficiently small $e/m$ exhibits a nonvanishing electric field $\langle \hat E(x) \rangle\neq 0$, which is inhomogeneous due to the finite system size. 
Turning off the bias field, $\theta\rightarrow 0$, which corresponds to a quantum quench as indicated in Fig.~\ref{fig:QED_illustration}a, induces a discharge of the capacitor as dynamical charges are produced and accelerated across the capacitor~\cite{gold2021backreaction}, thus reducing the background electric field, as sketched in Fig.~\ref{fig:QED_illustration}b (see ~\cite{zache2019dynamical} for another motivation to study quenches of the $\theta$ angle).
The dynamical processes are reminiscent of Schwinger pair production and string breaking in strong fields, which leads to plasma oscillations due to the coupling between the gauge field and the fermionic matter~\cite{hebenstreit2013simulating}.
In the dual picture, these processes can be probed by monitoring the damped oscillations of $\langle \hat E \rangle = -e\langle \hat \varphi \rangle / (2\pi)$ together with the charge density \mbox{$\langle \hat \rho \rangle = -e\langle \partial_x \hat \varphi \rangle/(2\pi)$}.
In general, studying these dynamical processes is computationally hard, which motivates us to address this problem with an analog quantum simulator. 

While an experimental realization of our proposed setup will be able to faithfully simulate the full dynamics in the massive SG model, we necessarily restrict ourselves to a relatively small lattice system with small spin lengths to illustrate the qualitative behaviour in a numerical simulation.
To be specific, we consider a lattice with $N = 7$ atoms, spin length $J = 4$ and $\kappa = 4$. 
The result of a quench $\theta = \pi/2 \rightarrow 0$ is shown in Figs.~\ref{fig:QED_illustration}c-d, where the chosen parameters are indicated in the caption. 

In Fig.~\ref{fig:QED_illustration}c we show the real-time dynamics of $\langle \hat J^{(i)}_y\rangle$ and the finite difference $\langle \hat J^{(i)}_y - \hat J^{(i-1)}_y\rangle$, which correspond to the lattice version of the electric field $\langle \hat E_i\rangle$ and the charge density $\langle \hat \rho_i\rangle$, respectively, up to appropriate proportionality constants.
Even for the small system size and finite spin length considered here, we observe the expected damped oscillations of the electric field and the corresponding charge dynamics.
These are driven by the interplay of pair creation and the subsequent interaction among the fermions with the gauge fields. 
At later times, we find partial revivals that we attribute to finite-size effects, thus prohibiting us from studying the long-time limit in our simulations. 
Nevertheless, as shown in Fig.~\ref{fig:QED_illustration}d, the plasma oscillations show an inhomogeneous damping, consistent with the expectation that the electric field decays in the middle of the capacitor~\cite{gold2021backreaction}, while a finite charge density accumulates at the system's boundaries, leading to an effective screening effect, reminiscent of string breaking.
In contrast to our benchmark simulations, an experimental realization can be scaled up to large spin lengths $J\gg 1$ and large lattice sizes, namely $N\gg1$ Rydberg atoms, with the perspective to reach the QFT limit. Such a realization will enable a quantitative analysis of the intricate behaviour towards later times, which can help to improve our general understanding of the thermalization process in gauge theories~\cite{berges2021qcd,mueller2021thermalization}.

\section{Quantum information prospects}
\label{sec:ProspectQI}

\begin{figure}[!t]
\center
\includegraphics[width=1\columnwidth]{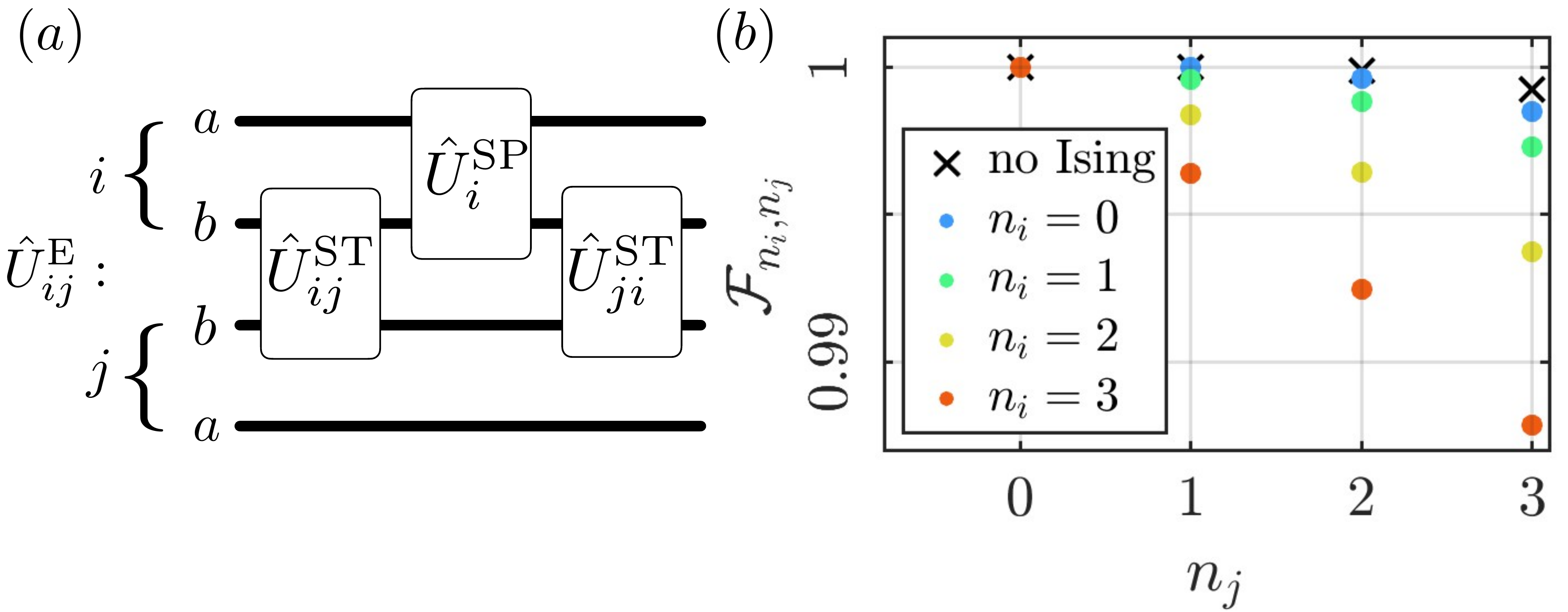}
\caption{Entanglement generation and state transfer protocol. (a) Gate sequence to perform an entangling gate $\hat U^{\mr E}_{ij}=\hat U^\mr{ST}_{ij} \hat U_i^{\mr{SP}}  \hat U^\mr{ST}_{ji}$ on two Rydberg atoms for states near the circular level. (b) Fidelity of the state transfer protocol $\hat U^\mr{ST}_{ij}$ for different initial states 
obtained with the spin Hamiltonian including flip-flop and Ising terms (dots) and without the Ising terms (crosses) with $J=50$.} 
\label{fig:StateTransfer}
\end{figure}

The large and regularly structured Rydberg manifold together with the high level of external control and manipulation available for such a system offer an opportunity to exploit this platform for qudit-based quantum information processing.
As an illustration we present an entangling gate $\hat U_{ij}^\mr{E}$ between a pair of atoms $i$ and $j$, where the quantum information of atom $i$ ($j$) is stored in its $\mathcal{H}^a$ ($\mathcal{H}^b$ defined in the beginning of Sec.~\ref{subsec:edge}) manifold as depicted in Fig.~\ref{fig:StateTransfer}a.
A state transfer operation $\hat U_{ij}^\mr{ST}$ moves the quantum information from the $b$-degree of freedom of atom $j$ to the $b$-degree of freedom of atom $i$, where a general single particle operation $\hat U_{i}^\mr{SP}$ acting on atom $i$ subsequently entangles the two subspaces, before transferring the $b$ degree of freedom back to atom $j$. 

We now discuss the state transfer gate $\hat U^\text{ST}_{ij}$ that is inspired by the one developed for coupled harmonic oscillators, see for example \cite{Portes2008}.
The state transfer protocol can be realized near the circular level $\ket{0,0}$ by exploiting the dipole-dipole interactions, see Sec.~\ref{subsubsec:NearCircularState}, as we show at the end of this Section.
To this end, consider the initial two-atom state
\mbox{$\ket{\psi_0} = \ket{n_i, 0}_i\otimes \ket{0, n_j}_j$}, where the atom $i$ encodes information in the $\mathcal H^a$ manifold and the atom $j$ encodes information in the $\mathcal H^b$ manifold.
In order to transfer the information from atom $j$ to atom $i$, we consider the Hamiltonian $\hat H_{ij}^{\mr{ST}} \!=\!- h_{ij} \hat c_{i b}^\dagger \hat c_{j b}^{}\!+\!\mr{H.c.}$, see Eq.~\eqref{eq:HPHamiltonian}, valid for $n_i,n_j \ll J$,  that generates the unitary $\hat U_{ij}^{\mr{ST}} \!=\!\exp\! \big(\!-\!i\hat H_{ij}^{\mr{ST}}T\big)$. Taking $T = \pi/(2h_{ij})$, the application of the unitary yields
\begin{align}
    \ket{\psi_T} \equiv \hat U_{ij}^{\mr{ST}} \ket{\psi_0} \!=\! (-i)^{n_j} \ket{n_i, n_j}_i\otimes \ket{0, 0}_j,\notag
\end{align}
which 
achieves the desired state transfer, see App.~\ref{Appendix:StateTransfer} for further details. Generalization to arbitrary initial superposition states is straightforward. 

Entanglement between the two atoms can then be generated by constructing a more general entangling gate $\hat U^\mr{E}_{ij}$ defined by the sequence 
\begin{align}
    \hat U^\mr{E}_{ij} = \hat U^\mr{ST}_{ij} \hat U_i^{\mr{SP}}  \hat U^\mr{ST}_{ji}\,,
\end{align}
see Fig.~\ref{fig:StateTransfer}a, where the single-particle unitary $\hat U^\mr{SP}_{i}$ required for the entanglement generation between the two atoms can be realized, for example, by the ponderomotive manipulation techniques discussed in Sec.~\ref{subsubsec:Pomderomotive}.

The implementation of $\hat U^\mr{ST}_{ij}$ relies on the dipole-dipole interaction Hamiltonian, which we approximated with $\hat H_{ij}^{\mr{ST}}$.
From inspecting the effective dipole-dipole interaction Hamiltonian, Eq.~\eqref{eq:HPHamiltonian}, we see that undesired pair-creation terms can be eliminated by tuning $|\omega_a - \omega_b|\gg JV_{ij}$. 
Furthermore, we can isolate the $b$ hopping (or flip-flop) terms, by using electric and magnetic gradient fields that shift the $a$ hopping processes out of resonance by making $\omega_a$ spatially dependent.
Finally, density-density (or Ising) interactions can be neglected in the limit $n_i,n_j \ll J$.

In Fig.~\ref{fig:StateTransfer}b, we quantify the accuracy of the state transfer protocol by plotting the gate fidelity $\mathcal F_{n_i,n_j} = \bra{\psi_T}\exp(-i\hat H^t T)\ket{\psi_0}$. 
Here we take several possible initial states
and we include only Ising and $b$ flip-flop terms in $\hat H^t$.
We find that a high fidelity $\mathcal F_{n_i,n_j} \gtrsim 0.99$ can be obtained for $n_i,n_j \lesssim 3$ and $J=50$.
The main source of error is originating from the finite spin length, which affects both the Ising and the flip-flop terms via the Holstein-Primakoff transformation. 
The fidelity may be further improved by using optimal control techniques \cite{Calarco2015} or by adapting the scheme demonstrated in Refs.~\cite{Geier2021,Browaeys2022} to our system that could suppress the effect of the Ising terms.

%%%%%%%%%%%%%%%% Conclusion   %%%%%%%%%%%%%%%%
\section{Conclusions}
\label{sec:conclusions}

In this work, we have presented an hardware-efficient quantum simulation Rydberg toolbox for atoms in tweezer arrays based on the high-dimensional manifold of states with principal quantum number $n$ in the regime of linear Stark and Zeeman effect.
We have exploited the SO(4) symmetry of the hydrogen-like states to characterize these energy levels in terms of two angular momenta of large length, which have been conveniently used to represent the action of external fields (static electric and magnetic fields as well as optical and microwave fields).
Within this formalism, we have shown that the many-body problem is represented by generalized `large-spin' Heisenberg models whose large local Hilbert space can be used to encode conjugate (continuous) variables to simulate QFTs.
In particular, we have illustrated how to realize the 1D sine-Gordon model in the massless and massive case.
Besides the quantum simulation applications, these high-dimensional manifolds and their control via external fields can also be exploited for qudit-based quantum information processing, which we have exemplified with an entangling gate and a state-transfer protocol involving states in the vicinity of the circular level.

Our results offer the opportunity to simulate more general QFTs in higher dimensions, gauge degrees of freedom with large occupation numbers or large-spin models in arbitrary lattice geometries.
As our analysis has focused on encoding a local Hilbert space on a single $n$-manifold, a possible extension is to study models where multiple manifolds are considered or where the electron spin is also included. 
Furthermore, other possible future directions include to employ these Rydberg states to encode synthetic dimensions \cite{Ozawa2019} or to achieve quantum-enhanced sensing \cite{kitagawa1993squeezed,Evrard2019}.

\section*{Acknowledgements}
The authors thank M.~Baranov, F.~Ferlaino, R.~Kaubr\"ugger, A.~Kaufman, C.~Kokail,  M.~Mark, and A.~M.~Rey for valuable discussions. This work was supported by
the US Air Force Office of Scientific Research (AFOSR)
via IOE Grant No. FA9550-19-1-7044 LASCEM, the European Union's Horizon 2020 research and innovation program under Grant Agreement No. 817482 (PASQuanS), the \mbox{QuantERA} grant MAQS via the Austrian Science Fund FWF No I4391-N,
and the Simons Collaboration on Ultra-Quantum Matter, which is a grant from the Simons Foundation (651440,
P.Z.).

%%%%%%%%%%%%%%%% Appendix   %%%%%%%%%%%%%%%%

\appendix
\section*{Appendix}

\section{Algebraic solution of the Hydrogen problem}
\label{sec:AppendixHydrogen}

Here, we summarise further details of the algebraic solution of the hydrogen atom outlined in the main text Sec.~\ref{sec:algebraicsolution}. 

As already pointed out in the main text, the non-relativistic Hydrogen Hamiltonian reads $\hat H_\mathrm{H} = {\hat{\mb{p}}^2}/{2} - {e}/{4\pi\epsilon_0|\hat{\mb r}|}$, and a convenient approach to solve for the eigenvalues is to use constants of motion.
In particular, the  the orbital angular momentum $\hatboldL = \hat{\mb r} \times \hat{\mb p}$ and the Runge-Lenz (RL) vector
\begin{align}
\hat{\mb A}_0 = \hat{\mb r}/|\hat{\mb r}|-4\pi\epsilon_0(\hat{\mb p}\times \hatboldL - \hatboldL\times \hat{\mb p})/(2m_r e^2)    
\end{align}
are commuting with $\hat H_\mr{H}$.
At the classical level, the RL vector is a conserved quantity of the Kepler problem, it is aligned with the major axis of a closed orbit and its magnitude is proportional to its eccentricity.
Therefore, the RL vector is orthogonal to the orbital angular momentum vector, which is also inherited by the quantum problem, namely $\hatboldL\cdot \hat{\mb A}_0 = 0$, and will be used below.

As we are interested in bound state solutions ($E<0$), it is useful to redefine the RL vector 
\begin{align}
\hat{\mb A} = \sqrt{-\mr{Ry}/\hat H_\mathrm{H}}\hat{\mb A}_0,
\end{align}
such that we obtain commutation relations of the form $[\hat{L}_{i}, \hat{L}_{j}]=i\epsilon_{ijk} \hat{L}_k$, $[\hat{A}_{i}, \hat{A}_{j}]=i\epsilon_{ijk} \hat{L}_k$ and $[\hat{L}_i,  \hat{A}_j] = i\epsilon_{ijk}\hat{A}_{k}$ that form the SO(4) Lie algebra.
It is convenient to change the basis of the algebra to SU(2)$\times$SU(2) 
\begin{align}
\hat{\mb J}_a = \frac{1}{2}\left( \hatboldL -\hat{\mb A}\right)~\mathrm{and}~ \hat{\mb J}_b = \frac{1}{2}\left( \hatboldL +\hat{\mb A}\right)\,,
\label{eq:JaJbApp}
\end{align}
which doubly covers SO(4). The commutation relations of the two angular momenta $\hat{\mb{J}}_a$ and $\hat{\mb{J}}_b$ are $[\hat J_{\sigma,i},\hat J_{\sigma',j}] = i\delta_{\sigma,\sigma'}\epsilon_{ijk} \hat J_{\sigma,k}$, where $\sigma, \sigma' = a, b$.

From the nontrivial relation \mbox{$\hat H_\mr{H}(\hatboldL{}^2+ \hat{\mb A}^2+1) = -\mr{Ry}$}, it is possible to rewrite the Hydrogen atom Hamiltonian as $\hat H_\mathrm{H} = -\mr{Ry}/[2(\hat{\mb J}_a^2 +  \hat{\mb J}_b^2)+1]$. 
Furthermore, the orthogonality between the orbital angular momentum and the RL vector ($\hatboldL\cdot \hat{\mb A} = 0$) also fixes the length of the two angular momenta,  $\hat{\mb J}_a^2 = \hat{\mb J}_b^2 = (\hatboldL{}^2 + \hat{\mb A}^2)/4$, such that an obvious basis for the hydrogen atom eigenstates is $\ket{J,m_a,m_b}$, where
\begin{align}
\hat{\mb{J}}_{a\,(b)}^2\ket{J,m_a,m_b} &= J(J+1)\ket{J,m_a,m_b}~\mr{and}\notag\\
\hat{J}_{a\,(b),z} \ket{J,m_a,m_b} &= m_{a\,(b)}\ket{J,m_a,m_b},
\end{align}
with $J = 0,1/2,1,\dots$ and $m_{a\,(b)}\in\{-J,\,-J+1,\,\dots,\,J\}$. 
The corresponding energies of the Hamiltonian take the form $E_n = -\mr{Ry}/(2J+1)^2 = -\mr{Ry}/n^2$ from which we can read off the relation between the principal quantum number $n$ and the angular momentum length $J = (n-1)/2$. 
For typical Rydberg states where $n\gg 1$, one therefore obtains two large angular momenta with $J \gg 1$ and an enormous $n^2 = (2J + 1)^2$ degeneracy of the energy levels.

From the nontrivial commutator 
\begin{align}
i[\hat H_\mr{H}/\mr{Ry},\,\hatboldL\times \hat{\mb r}+|\hat{\mb r}|^2\hat{\mb p}] = 6\hat{\mb A}_0a_0 + 4\hat{\mb{r}} \hat H_\mr{H}/\mr{Ry},
\end{align}
it is straightforward to notice that the matrix elements of the left-hand side of this equation within a specific $J$ manifold vanish because they are all eigenstates of $\hat H_\mr{H}$ with the same energy $E_n$.
As a result, we obtain that the eletric dipole operator $\hat{\bs{\mu}} = -e \hat{\mb r}$ projected on a specific $n$ manifold takes the simple form
\begin{align}
\hat{\bs{\mu}} = -\frac{3\,ea_0\,n}{2}\hat{\mb{A}}=\frac{3\,ea_0\,n}{2}\big(\hat{\mb J}_a - \hat{\mb J}_b\big)\,,
\label{eq:dipApp}
\end{align}
see Refs.~\cite{Flamand1966, Becker1976, Valent2003} for further details of the derivation.

\section{Decoherence}
\label{Appendix:Decoherence}
In this Appendix we discuss decoherence channels of the 
Rydberg levels discussed in the main text Sec.~\ref{sec:generalization}. In particular, we present spontaneous emission rates and discuss motional decoherence.

\begin{figure}[!t]
\center
\includegraphics[width=1.0\columnwidth]{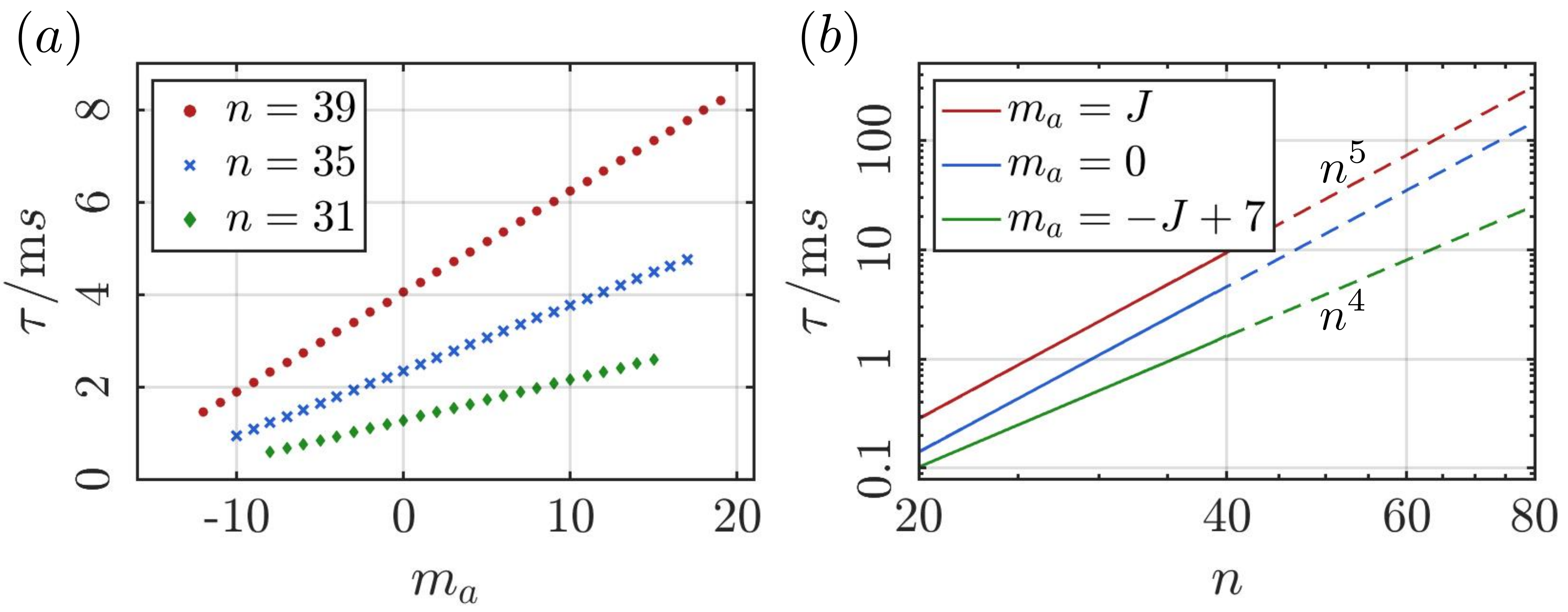}
\caption{Lifetime: (a) Lifetime of the levels $\ket{\psi^a_{m_a}}\in\mathcal{H}^a$ for different principal quantum numbers. Note, the lifetime for states with $J+m_a<7$ is out reach for the numerical methods, as the radiative decay to low lying states requires knowledge of the ground- and low lying excited state wavefunctions. 
(b) Scaling of the lifetime of the states $\ket{\psi^a_{m_a}}$ with $n$. The solid lines are numerically calculated lifetimes and the dashed lines are a fit of the monomial $n^\alpha$, where $\alpha = 5$ for $m_a = J,0$ and $\alpha = 4$ for $m_a = -J+7$.
} 
\label{fig:SpontaneousEmissionAppendix}
\end{figure}

\subsection{Spontaneous emission}
\label{Appendix:SpontaneousEmission}

In a cryogenic environment (i.e. for temperatures below a few K \cite{Nguyen2018}) the lifetime of the Rydberg levels is fundamentally limited by spontaneous emission to lower lying levels.     
The decay rate from an initial state $\ket{i}$ with energy $\omega_i$ to a final state $\ket{f}$ with energy $\omega_f$ is determined by the spontaneous emission rates 
$\gamma_{i, f} = \omega_{i,f}^3|\bra{i}\hat{\bs{\mu}}\ket{f}|^2/(3\pi\epsilon_0 c^3)$ \cite{carmichael1993open},
where $\omega_{i,f} = \omega_i- \omega_f$ is the transition frequency and $\bra{i}\hat{\bs{\mu}}\ket{f}$ the dipole transition matrix element. 
The total lifetime $\tau_i$ of $\ket{i}$ is then given by $\tau^{-1}_{i} = \sum_{f<i} \gamma_{i, f}$, where the $\sum_{f<i}$ is limited to final states with $\omega_f < \omega_i$.

In Fig.~\ref{fig:SpontaneousEmissionAppendix}a we present the numerically calculated lifetimes for the manifold $\mathcal{H}^a$ (introduced in the main text Sec.~\ref{sec:manybodymodels}) for different principal quantum numbers $n$.
The relatively large lifetime of the circular level ($m_a = (n-1)/2$) is to a large extent preserved for states with lower orbital angular momentum components.
In Fig.~\ref{fig:SpontaneousEmissionAppendix}b we monitor the scaling of the lifetime with $n$. 
The lifetime scaling of the circular level is $n^5$ \cite{Nguyen2018} and it gradually changes to the numerically calculated $n^4$ scaling for states with low orbital angular momentum components. 
The $n^4$ lifetime scaling of states with low angular momentum components can be understood qualitatively as follows: the lifetime of low orbital angular momentum states typically scales as $n^3$ \cite{Low2012}; if these states are not affected by quantum defects, in the presence of an external electric field they are  distributed among $O(n)$ Stark states \cite{Biedenharn1984} and, therefore, their lifetime is enhanced by a factor $n$.

\subsection{Motional decoherence}
\label{Appendix:MotionalDecoherence}
Here we summarize decoherence mechanisms associated with motional effects. 
First we discuss magic (state independent) trapping of the Rydberg states considered in this paper and then estimate motional excitation rates due to dipole-dipole forces. 

It has been proposed \cite{Dutta2000, Cohen2021}  and experimentally demonstrated \cite{Anderson2011, Barredo2020, Cortinas2020} that Rydberg states can be spatially confined using ponderomotive trapping techniques. 
For hollow bottle beam tweezers the size of the Rydberg wavefunctions is typically comparable to the radial extent of the tweezer trap and, hence, the ponderomotive potential typically depends on the Rydberg state.
Therefore, finding magical trapping conditions that avoid internal state dephasing is challenging with ponderomotive traps.

An alternatively route available for Alkaline Earth and Lanthanide atoms is to use the polarizability of the optically active core to trap the atom with a red-detuned Gaussian tweezer beam \cite{Topcu2014, Mukherjee2011, Wilson2022}. 
In particular, when the trapping light is almost resonant with a core transition, the core polarizability becomes much greater than the ponderomotive potential (from the trapping light) experienced by the Rydberg electron.
Therefore the trapping potential becomes to a large extent independent of the Rydberg state \cite{Cohen2021,Wilson2022}, which enables magic trapping for many different Rydberg states.

Next, we estimate the creation of vibrational excitations due to dipole-dipole forces acting on the nucleus inside a harmonic trap with frequency $\omega_t$.
An interaction with effective strength $V_{ij}J(J+1)$, as considered in Sec.~\ref{sec:SGmodel} of the main text, gives rise to an additional potential $U( R_i) =  3R_iV_{ij}J(J+1)/|R_{ij}|$ acting on the $i$-th atom, where $ R_i$ denotes the position of the $i$-th nucleus measured from the center of the trap. 
If the strength of $U(R_i)$ at the characteristic trapping length scale $l_h= \sqrt{1/(M_a\omega_t)}$ ($M_a$ is the mass of the atom) is small compared to $\omega_t$, \emph{i.e.} $U(l_t)\ll \omega_t$, then at short times ($\omega_t t/(2\pi)\ll 1$) excitations are created as $n(t) = U(l_t)^2 t^2/8$, where we assumed that the nucleus is initialized in the ground state of the oscillator, which can be achieved by sideband cooling  \cite{Kaufman2022,Ma2022}.

As an example we consider the parameters used to realize the SG model from Sec.~\ref{sec:SGmodel} of the main text, for which we have an interaction strength of $V_{ij}J(J+1)= 2\pi\times 300\,\mr{kHz}$ for $n = 41$ and an inter-atomic separation of $R_{ij} = 17\mu\mr{m}$. 
Assuming a typical trapping frequency of $\omega_t = 2\pi\times40\,\mr{kHz}$ and strontium atoms ($M_a \approx 87 \mr{amu}$) gives rise to $l_t = 54\,\mr{nm}$ for which we get  $U(l_t) \approx 2\pi\times 3 \, \mr{kHz}\ll\omega_t$.
For these parameter values, $n(10t_c) \approx 0.05$ excitations are created after 10 interaction cycles of duration $t_c = 2\pi/[V_{ij}J(J+1)]$.

\section{Ponderomotive coupling}
\label{subsec:ponderomotive}

In the following we discuss the ponderomotive manipulation techniques introduced in Sec.~\ref{subsec:triangular} of the main text in more detail. The discussion is structured as follows: First we introduce Laguerre-Gauss (LG) laser beams and then we discuss the ponderomotive potential generated by interfering two LG beams.
\newline

\textit{Laguerre-Gauss laser beam:} The vector potential of a linearly polarized LG laser beam with orbital angular momentum $\delta m$ and mode number $p$, propagating along the $z$-axis, is in the Lorenz gauge given by \cite{Kogelnik1966,Kimel1993}
\begin{align}
    \mb{A}_{p,\delta m}(\mb r,\, t) =&\, \hat{\mb{e}}_x\, A_{p,\delta m} \, u_{p,\delta m}(\mb r)e^{i\alpha}\notag\\
    &\times\,e^{-i\delta m\phi}\,e^{-i(kz-\omega t)}/2 + \mathrm{c.c.},
    \label{eq:LGbeam}
\end{align}
where we used cylindrical coordinates ($\rho = \sqrt{x^2+y^2}$ and $\tan \phi  = y/x$).
Here $\omega$ is the light frequency, $k = 2\pi/\lambda$ is the magnitude of the corresponding wave vector, $\lambda$ is the wavelength and $\alpha$ is a phase.  
Moreover, $A_{p,\delta m}$ denotes the field amplitude and $u_{p,\delta m}(\mb r)$ is a dimensionless field distribution function satisfying the Helmholtz equation. In the paraxial limit $u_{p,\delta m}(\mb r)$ can be expressed in terms of associated Laguerre polynomials $L_p^{(|\delta m|)}(x)$ as
\begin{align}
    u_{p,\delta m}&(\rho, z) =\, C_{p,\delta m}\,\frac{w_0}{w(z)}\,\bigg(\frac{\sqrt{2}\rho}{w(z)} \bigg)^{|\delta m|}\notag\\
    &\times L_p^{(|\delta m|)} \bigg(\frac{2\rho^2}{w(z)^2} \bigg) \exp\bigg(\frac{-\rho^2}{w(z)^2} \bigg)\\
    &\times \exp \bigg(i\bigg[\frac{k\rho^2z}{2(z^2+z_R^2)}-(2p+|\delta m|+1)\mr{atan}\frac{z}{z_R} \bigg] \bigg)\notag.
    \label{eq:LGbeamdistribution}
\end{align}
 The beam waist is defined as $w(z) = w_0\sqrt{1+z^2/z^2_R}$, where $w_0$ denotes the waist in the focal plane and $z_R = \pi w_0^2/\lambda$ is the Rayleigh length. The normalization constant $C_{p,\delta m}= \sqrt{p!\,2/(p+|\delta m|)!}$ is chosen such that $\int\,\mr{d}A |u(\rho,z)|^2 = w_0^2 \pi$. Due to this normalization the time averaged electric beam power $P = c\epsilon_0\omega^2|A_{p,\delta m}|^2w_0^2\pi/4$ only depends on the field amplitude and not on the LG mode numbers. 
\newline

\textit{Ponderomotive potential:} The ponderomotive potential that a highly excited Rydberg electron experiences from a fast oscillating laser field is given by \cite{Bucksbaum1988} \mbox{$U_\mr{PM}(\mb R+\mb{r},t) =e^2|\mb{A}(\mb{R}+\mb{r},t)|^2/(2m_e)$}, where $e$ and $m_e$ are the electron charge and mass, respectively. 
The vector potential of the laser field is evaluated at the position of the electron in the laboratory frame, \emph{i.e.} $\mb{R}$ is the position of the core and $\mb{r}$ is the position of the electron relative to the core. 
In the tight trapping limit, when the size of the vibrational core wave-packet is much smaller than the LG waist $w_0$, $\mb{R}$ can be replaced by its mean value. 
Furthermore, if the laser beam is focused onto the center of the trap, \emph{i.e.} $\langle \mb{R}\rangle = 0$,  which can be achieved by re-using the trapping light optics, the ponderomotive potential becomes approximately independent of the core position $U_\mr{PM}(\mb R+\mb{r},t)\approx U_\mr{PM}(\mb{r},t)$.

As already mentioned in the main text, the core idea to couple states with multiple orbital angular momentum $\kappa\geq 1$ difference is to use  the ponderomotive potential of two co-propagating LG laser beams with different oscillation frequencies and non-zero orbital angular momentum \cite{Dutta2000,Knuffman2007, cardman2020circularizing, Cohen2021}. 
Using Eq.~\eqref{eq:LGbeam}, the vector potential of the two co-propagating LG beams is given by
\begin{align}
    \mathbf{A}_\mr{LG}&(\mb r,\, t) =\, \hat{\mathbf{e}}_x\, A_\mr{LG} \big[
    u_{\delta m_1}(\rho,z)e^{-i\delta m_1 \phi}e^{-i(\omega_1 t - k_1 z-\alpha_1)}\notag\\
    +&u_{\delta m_2}(\rho,z)e^{i\delta m_2 \phi}e^{-i(\omega_2 t-k_2 z-\alpha_2)} +
    \mathrm{c.c.}\big]/2,
\end{align}
where the subscripts $1$ and $2$ are labels for the first and second LG beam.
The field distribution functions $u_{\delta m_{1(2)}}$ are in the most general case a superposition of multiple LG modes $u_{p,\delta m_{1(2)}}$ from Eq.~\eqref{eq:LGbeamdistribution}, with fixed $\delta m_{1(2)}$.
For simplicity, the field amplitudes are assumed to be the same for both beams. 

The corresponding ponderomotive potential, after dropping fast oscillating terms, becomes 
\begin{align}
    U_\mr{PM}(\mb{r},t) \approx \big[ U_c(\mb r) + 
    (U_\kappa(\mb r)e^{-i\kappa\phi}e^{-i\delta\omega t} + \mr{c.c.})\big],
    \label{eq:PMphase}
\end{align}
where $\delta\omega  = \omega_1 - \omega_2$ is the oscillation frequency of the interference term and $\kappa = \delta m_1 + \delta m_2$ is the transferred orbital AM. The spatial shapes of the constant and oscillatory terms are given by 
\begin{align}
U_c(\mb r) &= U_0\,\big[|u_{\delta m_1}(\rho,z)|^2 + |u_{\delta m_2}(\rho,z)|^2\big]/2\notag\\
U_\kappa(\mb r) &= U_0e^{i(\alpha_1-\alpha_2)}\,u_{\delta m_1}(\rho,z)\,u^*_{\delta m_2}(\rho,z)/4,
\label{eq:PonderomotiveContributions}
\end{align}
with $U_0 = e^2P_\mr{tot}/(16\pi^3m_e\epsilon_0c^3)\,(\lambda^2/w_0^2)$, where
$P_\mr{tot}$ is the combined power of both beams. 
The contributions of the ponderomotive potential are twofold: the time independent term gives rise to a level dependent energy shift  and the time dependent term can couple states that differ by $\kappa$ orbital angular momenta. 

\begin{figure}[!t]
\center
\includegraphics[width=1.0\columnwidth]{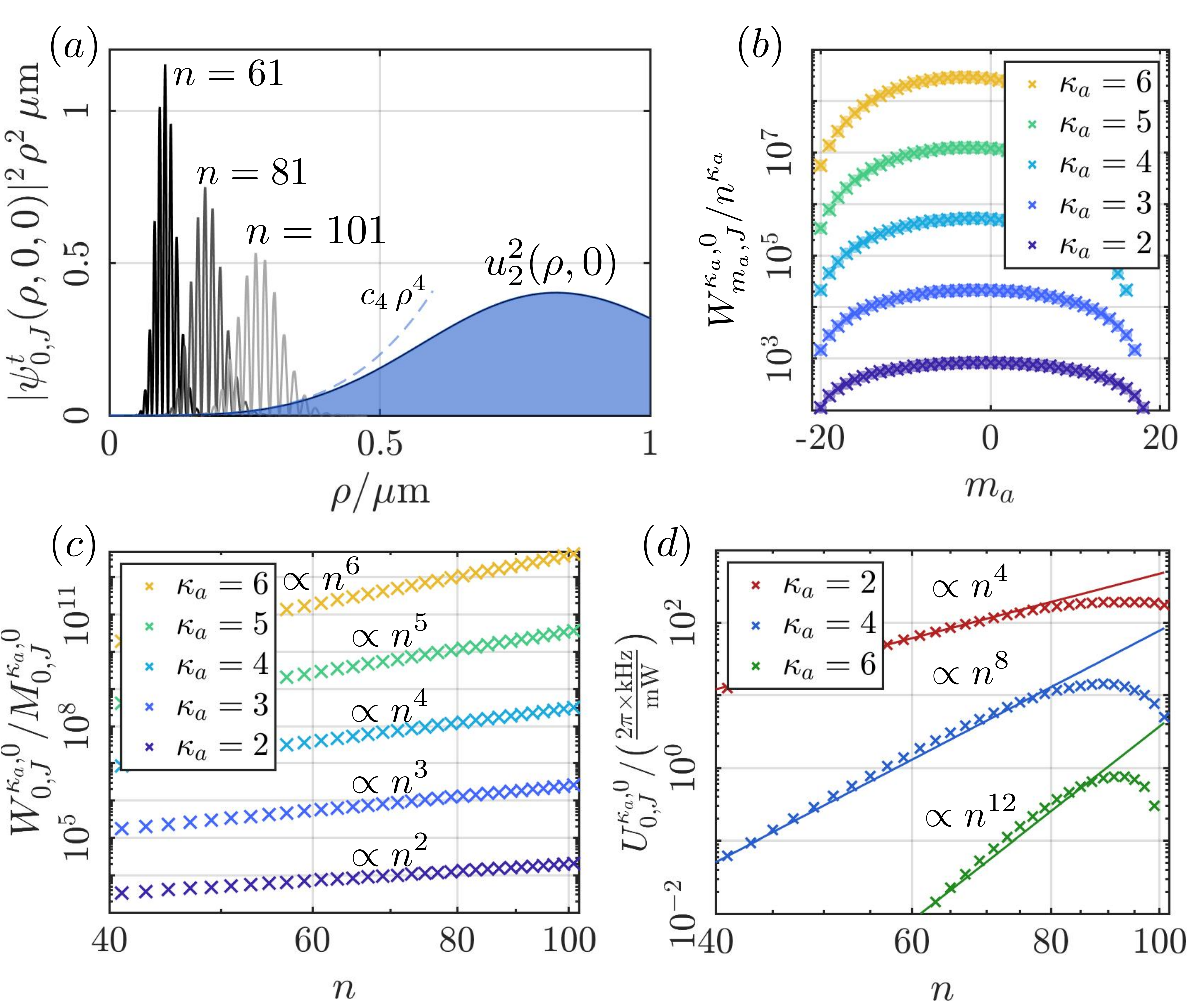}
\caption{Ponderomotive Coupling: (a) Illustrative comparison of the wavefunctions $\psi^t_{m_a,m_b}(\rho,\phi,z) = \langle \rho,\phi,z \ket{\psi^t_{m_a,m_b}}$ \cite{Biedenharn1984} and the spatial extent of the ponderomotive potential for $\delta m_{1(2)} = 2$, $\eta = 0$, $\lambda = 1300\,\mr{nm}$ and $w_0 = \lambda/2$. 
(b) Numerical verification of the relation Eq.~\eqref{eq:NumComparison} for $n=41$ and different values of $\kappa_a$. 
The numerically calculated matrix elements $W^{\kappa_a ,\kappa_b = 0}_{m_a,m_b=J}$ are displayed as crosses and the angular momentum matrix elements $C_{\kappa_a}M^{\kappa_a ,\kappa_b = 0}_{m_a,m_b=J}$ as circles, where $C_{\kappa_a}$ is a proportionality factor given by $C_{\kappa_a} = W^{\kappa_a ,0 }_{0,J} /M^{\kappa_a ,0 }_{0,J}$. 
The two sets of symbols lie perfectly on top of each other.
(c) Scaling of the proportionality factor. The crosses display $W^{\kappa_a ,0 }_{0,J}/M^{\kappa_a ,0 }_{0,J}$ for different values of $\kappa_a$ as a function of $n$. 
The scaling is given by $n^{\kappa_a}$, up to numerical errors due to a finite integration grid. 
(d) Numerical calculation of the transition matrix element $U_{m_a = 0,m_b = J}^{\kappa_a,\kappa_b=0}$ (crosses) for different principal quantum numbers. 
The solid lines are a fit of the monomial $n^{2\kappa_a}$ to $U_{ 0, J}^{\kappa_a,0}$.} 
\label{fig:PoderomotiveAppendix}
\end{figure}

Let us start by analyzing the time dependent contribution.
In particular, we are interested in the transition matrix elements of the form 
\begin{align}
U^{\kappa_a,\kappa_b}_{m_a ,m_b}=\bra{\psi^t_{m_a+\kappa_a, m_b+\kappa_b}}U_\kappa(\rho,z) e^{-i\kappa\phi} \ket{\psi^t_{m_a,m_b}}\;,
\label{eq:Ukappa}
\end{align}
that satisfy orbital angular momentum conservation \mbox{$\kappa_a+\kappa_b = \kappa$}. 
Note, $\kappa_{a(b)}$ can take on negative and positive values.
For the determination of the transition matrix element the functional behaviour of $U_\kappa(\rho,z)$ is of crucial importance. 
In the limit where the beam waist $w_0$ is greater than the typical extension of the Rydberg wave-function, see Fig.~\ref{fig:PoderomotiveAppendix}a, we can expand the ponderomotive potential in the focal plane $z=0$ around the phase vortex at $\rho = 0$, and obtain $U_\kappa(\rho,z=0)\approx   C_{\kappa+2\eta}\,\rho^{|\kappa|+2\eta}$, where the leading order $|\kappa| + 2\eta$, with $\eta\geq 0$, is determined by the LG beam modes. 

We now claim that, within the manifold $\mathcal{H}^t$, the matrix elements of the ponderomotive potential near the phase vortex $c_{\kappa+2\eta}\,\rho^{|\kappa|+2\eta}e^{-i\kappa\phi}$ are exactly proportional to those of $(\hat{J}_{a,+})^{\kappa_a} (\hat{J}_{b,+})^{\kappa_b}$, with $|\kappa_a| + |\kappa_b| =\kappa+ 2\eta$ and where we use the convention $(J_{a(b),+})^{-\kappa_{a(b)}} = (J_{a(b),-})^{\kappa_{a(b)}}$. More specifically, 
\begin{align}
W^{\kappa_a,\kappa_b}_{m_a,m_b} \!\!&\propto n^{|\kappa|+2\eta} M^{\kappa_a,\kappa_b}_{m_a,m_b},
\label{eq:NumComparison}
\end{align}
with
\begin{align}
W^{\kappa_a,\kappa_b}_{m_a,m_b} \!\!&=\! \bra{\psi^t_{m_a+\kappa_a,m_b+\kappa_b}}\!(\rho/a_0)^{|\kappa|+2\eta} e^{-i\kappa\phi}\! \ket{\psi^t_{m_a,m_b}}\!,\notag
\end{align}
and
\begin{align}
M^{\kappa_a,\kappa_b}_{m_a,m_b} \!\!&=\! \bra{\psi^t_{m_a+\kappa_a,m_b+\kappa_b}}\!(\hat{J}_{a,+})^{\kappa_a} (\hat{J}_{b,+})^{\kappa_b}\! \ket{\psi^t_{m_a,m_b}}.\notag
\end{align}
As it is quite tedious to find an exact algebraic relation, we instead confirmed the relation \eqref{eq:NumComparison} numerically and found perfect agreement.
In Fig.~\ref{fig:PoderomotiveAppendix}b we show such a numerical comparison for the special case of $\kappa_b = 0$. 
Fig.~\ref{fig:PoderomotiveAppendix}c showcases the scaling with $n$.
Note, the relation \eqref{eq:NumComparison} was inspired by the findings of Ref.~\cite{Becker1976}, who derived similar results for the matrix elements of $z^2$.

To analyse the ponderomotive transition matrix elements $U^{\kappa_a,\kappa_b}_{m_a ,m_b}$ from Eq.~\eqref{eq:Ukappa} it is important to take into account the full shape of $U_\kappa(\rho,z)$. 
As the functional behavior of $U_\kappa(\rho,z)$ is determined by the LG beam modes, it is sufficient to analyze $u_{\delta m_1}(\rho,0)\,u^*_{\delta m_2}(\rho,0)$.
A series expansion of the field distribution functions  at $z=0$ and around $\rho = 0$ yields in the most general case \mbox{$u_{\delta m_1}(\rho,0)\,u^*_{\delta m_2}(\rho,0) \! = \! c_{\kappa+2\eta}\rho^{|\kappa|+2\eta}\! + \! c_{\kappa+2\eta}^{(2)}\rho^{|\kappa|+2\eta + 2}+\dots\;$}. 
As only the leading term $c_{\kappa+2\eta}\rho^{|\kappa|+2\eta}$ generates the desired nonlinearity, higher orders have to be suppressed, which becomes essential for larger principal quantum numbers, as the increased size of the Rydberg-orbit starts to explore the full beam shape, see Fig.~\ref{fig:PoderomotiveAppendix}a. 

To suppress the higher orders we consider further beam shaping beyond a single LG beam mode, which can be achieved for instance by a spatial light modulator \cite{van2013quantum}.
Alternatively one could use super-positions of $N$ different LG modes, \emph{i.e.} $u_{\delta m_{1(2)}} = \sum^{N-1}_{p=0} a_{p}^{1(2)} u_{p,\delta m_{1(2)}}$, with $\sum (a_p^{1(2)})^2 = 1$, which allows to eliminate $2N-1$ higher orders $u_{\delta m_1}(\rho,0)\,u^*_{\delta m_2}(\rho,0)  = c_{\kappa+2\eta}\rho^{|\kappa|+2\eta}(1 + O(\rho^{2N}))$.
Hence, the scheme presented here becomes applicable for larger $n$ without increasing the beam waist $w_0$. 
However, with increasing $N$ the maximum of the ponderomotive potential moves further away from the vortex and, therefore, $U^{\kappa_a,\kappa_b}_{m_a ,m_b}$ for fixed beam power is reduced. 

For the transition matrix elements presented in Fig.~\ref{fig:MWPO}b in the main text, where we used the simplified notation $U_{\kappa_a,m_a,m_b} \equiv U^{\kappa_a,\kappa_b = 0}_{m_a,m_b}$, the LG beams are chosen as $u_{\delta m_1}(\rho, z) = u_{\delta m_2}(\rho, z)$, with $\delta m_1 = \delta m_2 = \kappa_a/2$ and $\eta = 0$. 
The number of modes used for $\kappa_a = 2$ and $4$ is $N = 2$ and for $\kappa_a=6$ it is $N=3$. 

For the same LG laser beams the transition matrix elements $U^{\kappa_a,\kappa_b = 0}_{m_a = 0,m_b=J}$ as a function of the $n$ are presented in Fig.~\ref{fig:PoderomotiveAppendix}d. 
For large principal quantum numbers the transition matrix elements deviate from the predicted $n^{2\kappa_a}$  scaling as the large Rydberg wavefunction probes the whole ponderomotive potential. 
Note, the scaling $n^{2\kappa_a}$ is obtained from Eq.~\eqref{eq:NumComparison} and the fact that the angular momentum transition matrix element $U^{\kappa_a,\kappa_b = 0}_{m_a = 0,m_b=J}$ scales for $n\rightarrow \infty$ as $n^{\kappa_a}$. 
We furthermore point out that by tuning the relative laser phase $\alpha_1-\alpha_2$ the coupling matrix elements $U^{\kappa_a,\kappa_b}_{m_a = 0,m_b}$ can be complex. 
Moreover, a pair of co-propagating LG beams generates a single nonlinear term and in order to drive the engineered transitions resonantly the laser frequencies have to satisfy $\delta \omega = -\kappa_a\omega_a + \kappa_b \omega_b$.

Let us briefly return to the first term of Eq.~\eqref{eq:PonderomotiveContributions}.
In contrast to the time dependent term, the constant part of the ponderomotive potential generates state dependent energy shifts. 
These shifts can in principle be used to engineer nonlinearities like higher powers of $J_{a,z}$ and $J_{b,z}$.
However, the strength of the shifts turns out to be relatively weak for reasonable laser powers of hundredths of mW per site, when compared to the nonlinearities generated by off-resonant microwave coupling.

Finally, we remark on decoherence effects induced by the LG beams due to  Thomson scattering of the nearly free Rydberg electron \cite{Cohen2021}. 
The scattering rate is given by $\Gamma_T = I\sigma_T/\omega$, where $I = 2P/(\pi\omega_0^2)$ is the mean beam intensity and $\sigma_T = (8\pi/3)e^4/(4\pi\epsilon_0m_ec^2)^2$ is the Thomson scattering cross-section \cite{Crowley2014} 
For $\lambda = 1300\,\mr{nm}$ and $w_0 = \lambda/2$ we obtain $\Gamma_T/P= 2\pi\times 0.7\,\mr{Hz}/\mr{mW}$, which is typically much less than the achieved coupling rates, see Fig.~\ref{fig:MWPO}b. Furthermore, we note that for the range of parameters considered here the Rydberg electron does not probe the high intensity regions (see Fig.~\ref{fig:PoderomotiveAppendix}a) and, therefore, the calculated scattering rates are overestimated.

\section{Offresonant MW coupling}
\label{Appendix:OffresonantMW}

\begin{figure}[!t]
\center
\includegraphics[width=1.0\columnwidth]{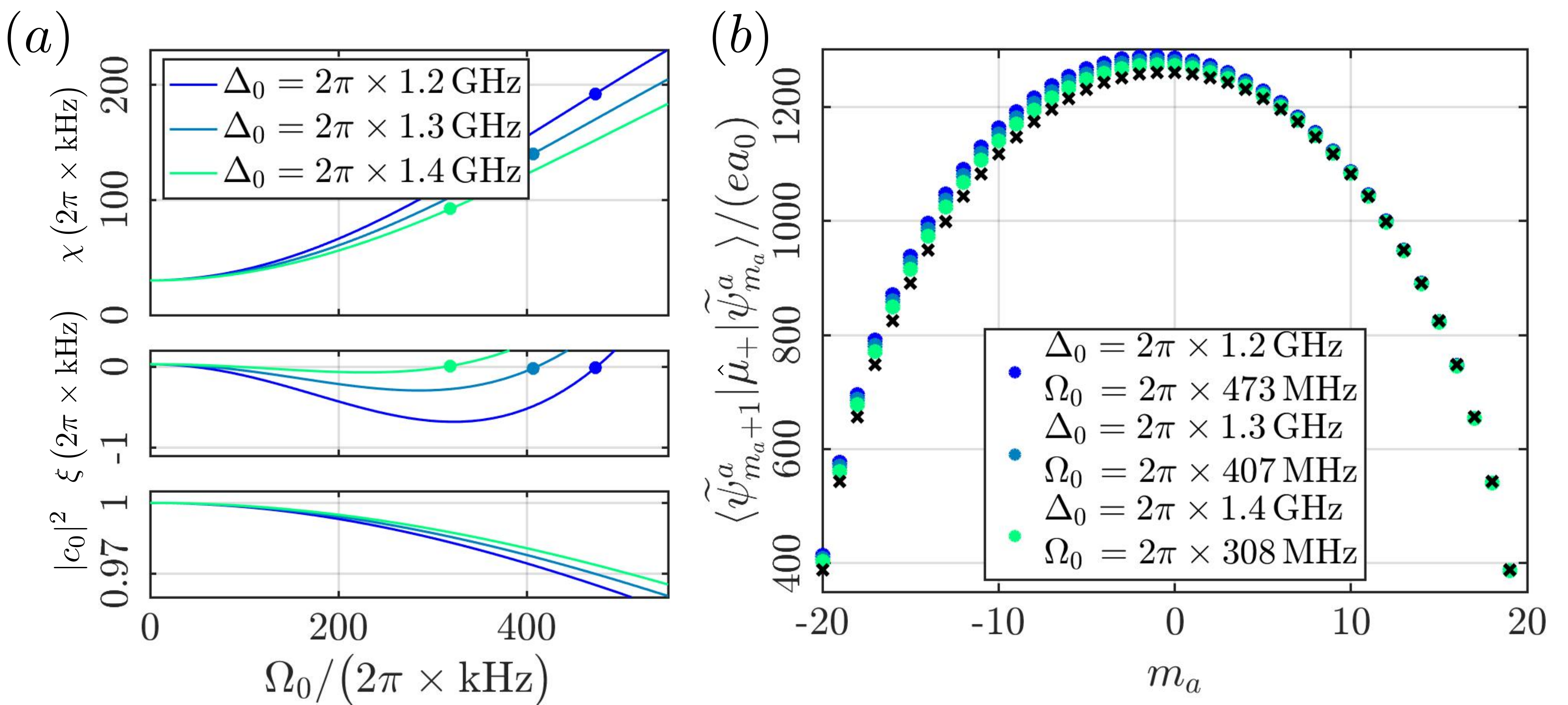}
\caption{Off-resonant MW coupling: (a) The top panel monitors $\chi$ for $n=41$ and different values of the MW detuning $\Delta_0$ and Rabi frequency $\Omega_{0}$. 
The electric field is $F = F_\mr{IT}/2$ and the magnetic field is chosen such that $\omega_a = -\omega_b/3$. 
The middle and lower panel present the strength of third order corrections $\xi$ and the state purity $|c_{m_a}|^2$ for $m_a = 0$, respectively.
The dots indicate the points where $\xi = 0$.
(b) Dipole transition matrix elements of the dressed states $\ket{\widetilde{\psi}_{m_a}^a}$, calculated for the MW parameters where $\xi = 0$.}
\label{fig:MWAppendix}
\end{figure}

This Appendix provides further details on the engineered Spin-squeezing term $\hat H_\mr{SQ}=\chi J_{z,a}^2$ introduced in the main text Eq.~\eqref{eq:Squeezing}.
To engineer this nonlinearity we consider $z$-polarized MW light that couples the manifold $\mathcal{H}_a$ with principal quantum number $n$ to the levels of the manifold $n' = n + \delta n$, with $\delta n > 0$. 
If the MW field is blue detuned with respect to the eigen-energies of the $n'$ manifold, only couplings to the manifold $\mathcal{H}_a$ of $n'$, i.e. the states $\ket{{\psi '}_{m_a}^{a}}$, are important, see main text Fig.~\ref{fig:MWPO}c. 
The underlying Hamiltonian is that of multiple independent two-level systems, one for each $m_a$, and is in a frame rotating at the MW frequency given by 
\begin{align}
    \hat H_\mr{SQ}^\mr{MW} =& -\sum_{m_a}\Delta_{m_a}\ket{{\psi '}_{m_a}^{a}} \bra{{\psi '}_{m_a}^{a}}\notag\\ 
    &+\sum_{m_a}\frac{\Omega_{m_a}}{2} \ket{{\psi '}_{m_a}^{a}}\bra{\psi_{m_a}^a} + \mr{H.c.},
\end{align}
with $m_a' = m_a - \delta n/2$. 
Here, $\Delta_{m_a} = \Delta_0-m_a \delta\omega_a$ is the level dependent detuning given in terms of the bare MW detuning $\Delta_0$ and the differential Stark shift $\delta \omega_a = \omega_a(n') - \omega_a(n) = 3\,ea_0\,\delta n F / 2$, see Fig.~\eqref{fig:MWPO}c. 
Moreover, $\Omega_{m_a} = 2 F_\mr{MW}\bra{{\psi '}_{m_a}^{a}}\hat \mu_z \ket{\psi_{m_a}^a}$ is the Rabi frequency, where $F_\mr{MW}$ is the electric field amplitude of the MW.

As already mentioned in the main text, in the far off-resonant limit ($\Delta_{m_a} \gg \Omega_{m_a}$), the dressed eigenstates $\ket{\widetilde{\psi}_{m_a}^a}$, which are adiabatically connected to $\ket{\psi_{m_a}^a}$ for $\Omega_{m_a}\rightarrow 0$, experience a state dependent AC-Stark shift $\Omega_{m_a}^2 / \big(4\Delta_{m_a}\big)$. 
Furthermore, in the limit $\Delta_0\gg m_a\delta\omega_a$ and due to the smooth dependence of $\Omega_{m_a}$ on $m_a$ the AC-Stark shift is expandable in powers of $m_a$,
\begin{align}
  \Omega_{m_a}^2 / \big(4\Delta_{m_a}\big) = \chi m_a^2+\xi m_a^3+O(m_a^4),
  \label{eq:MWexpansion}
\end{align}
where prefactors $\chi$ and $\xi$ are in the following determined numerically. 
Note, the constant and linear part are absorbed in $\Delta_\mr{MW}$ and $\delta \omega_a$, respectively.

The strength $\chi$ of the nonlinearity is limited by the electric field ionisation threshold, for which a lower bound is the Ingris Teller limit ($|F_\mr{MW}|<F_\mathrm{IT}$).
This limitation originates from the fact that the MW also couples to the permanent dipole moments $\bra{{\psi}_{m_a}^{a}}\hat \mu_z \ket{\psi_{m_a}^a}$ and thus induces a modulation of $\omega_{S}$, if $|F_\mr{MW}|$ exceeds $F_\mathrm{IT}$ the modulation leads to MW assisted ionisation \cite{Leopold1979}. 
For $|F_\mr{MW}| < F_\mathrm{IT}$, the modulation averages out because the MW frequency is much larger than the modulation amplitude $2 F_\mr{MW}\bra{{\psi }_{m_a}^{a}}\hat \mu_z \ket{\psi_{m_a}^a}$ for the range of parameters considered in this paper.
 
Within this limitation achievable prefactors $\chi$ can be on the order of hundreds of $2\pi\times\,$kHz, see Fig.~\ref{fig:MWAppendix}a upper panel. 
We point out that for every value of $\chi$ there exists a point in $\Delta_0$, $\Omega_{0}$ parameter space where the cubic contribution $\xi$ of Eq.~\eqref{eq:MWexpansion} vanishes, see Fig. \ref{fig:MWAppendix}a middle panel. 
Note, 2\textsuperscript{nd} order Stark corrections also give rise to nonlinearities (see Fig.~\ref{fig:MWAppendix}a upper panel $\Omega_0 = 0$), but these rapidly decrease with increasing $n$ as $1/n^{-6}$  \cite{Bethe1957, Becker1976}.
We also note that a misalignment of the MW polarization induces Raman transitions between neighbouring levels, which are however far off-resonant in the limit  $|\omega_a|\gg\Omega_{m_a}$.
As a result, $\chi$ is largely insensitive to polarization errors (below $10\%$ of $F_\mathrm{MW}$).

Due to the MW coupling the dressed eigen-states $\ket{\widetilde{\psi}_{m_a}^a}$ acquire character of the $n'$ manifold. The corresponding purity is quantified by $|c_{m_a}|^2 =|\langle {\psi}_{m_a}^a \ket{\widetilde{\psi}_{m_a}^a}|^2$ and decreases as $m_a\rightarrow -J$, since the effective MW detuning is reduced, see lower panel of Fig.~\ref{fig:MWAppendix}d. 
Furthermore, in the lowest panel of Fig.~\ref{fig:MWAppendix}a we monitor $|c_{m_a}|^2$ for $m_a = 0$ and different values of the detuning and driving strength.
The admixture of Rydberg levels from $n'$ to $\ket{\widetilde{\psi}_{m_a}^a}$ modifies the dipole transition matrix elements, which thus deviate from the angular momentum formalism introduced in Eq.~\eqref{eq:dip}. 
As an example of this deviation we present $\bra{\widetilde{\psi }_{m_a+1}^{a}}\hat \mu_+ \ket{\widetilde{\psi}_{m_a}^a}$, where $\hat \mu_+ = \hat\mu_x+i\hat\mu_y$ for different MW parameters in Fig.~\ref{fig:MWAppendix}b. 
The modifications of the dipole matrix elements are expected to be in practice negligible for the quantum simulation application presented in the main text Sec.~\ref{sec:SGmodel}.

\section{Exact results for the SG model}
\label{Appendix:SGExact}
For completeness, we provide exact expressions for the spectrum of the SG model in this Appendix, see e.g. \cite{Roy2021} and references therein. In the massive phase (\mbox{$\beta^2 <8\pi$}), there exist solitons with mass
\begin{align}
    M = \frac{2\Gamma \left(\frac{\xi}{2}\right)}{\sqrt{\pi} \Gamma \left(\frac{1+\xi}{2}\right)} \left[\frac{M_0^2(1+\xi) \Gamma \left( \frac{1}{1+\xi}\right)}{16\xi\Gamma \left(\frac{\xi}{1+\xi}\right)}\right]^{\frac{1+\xi}{2}}  \;,
\end{align}
where we abbreviated $\xi = \beta^2 / (8\pi - \beta^2)$.
For $\beta^2 < 4\pi$, the theory has additional bound states, called breathers. The number of different breathers depends the value of $\xi$ and they can be labelled by the integers $n =1, 2, \dots, \lfloor 1/\xi \rfloor$. The mass of the $n$-th breather is given by
\begin{align}
    m_n = 2M \sin \left(\frac{n\pi \xi}{2}\right)  \;.
\end{align}

\section{Implementing the sine-Gordon model with large spins}
\label{Appendix:SGimplementation}

In the following we outline how to simulate the SG model \eqref{eq:cont_SG} with the Rydberg simulator.
As pointed out in the main text, we start with a one-dimensional array of $N$ equally spaced Rydberg atoms. 
Isolating the subset of states $\mathcal{H}^a$ (see Sec.~\ref{sec:manybodymodels}), the system is described by the many-body Hamiltonian in Eq.~\eqref{eq:Hfull}.
Within the large spin limit ($J\rightarrow\infty$), we can employ the identification shown in  Eq.~\eqref{eq:CV_spin_identification} for every lattice site $i$. 
In order to obtain the desired limit $J\rightarrow \infty$, we rescale the interaction \mbox{$V_{ij}' = V_{ij} J(J+1)$}, while keeping $\Delta_a=0$, and $\lambda' = -2\lambda_\kappa [J(J+1)]^{\kappa/2}$.
For the special case $\kappa = 1$, this can be achieved by using a MW field as discussed before. 
The ponderomotive technique instead allows to achieve couplings with $\kappa > 1$, which gives access to a larger parameter regime of the SG model, as we discuss below. 
The resulting continuous variable lattice model reads
\begin{align}
    \label{eq:AppSGlat}
    \hat{H}_{\text{SG}}^{(\text{lat})} \!=\! &\sum_i \left[\chi \hat{\pi}_i^2 -  \lambda' \cos(\kappa\hat{\varphi}_i)  - \sum_{j> i}\frac{V_{ij}'}{2}\cos \left(\hat{\varphi}_{j}\!-\!\hat\varphi_i\right) \right] ,
\end{align}
where the Ising interactions are neglected in the limit $J\rightarrow\infty$.
The continuum SG model arises by keeping only the nearest-neighbor terms $V'_{i,i+1} \equiv V'_\text{nn}$.
To see this, we rescale the fields as $\kappa\hat{\varphi} \rightarrow \beta\hat{\varphi}$, $\hat{\pi}/\kappa \rightarrow (\ell/\beta)\hat{\pi}$ and the Hamiltonian as $\hat{H} \rightarrow (2\chi\kappa^2 \ell/\beta^2)\hat{H}$. 
Here, we introduced a short-distance scale $\ell$ that sets a UV-cutoff $\Lambda \propto 1/\ell$ \cite{jentsch2022physical}. 
As a consequence, we find that the lattice regularization $\hat{H}_\mr{SG}^{\mr{(lat)}}$ approximates $\hat{H}_\mr{SG}$ upon identifying the atomic couplings with SG parameters according to
\begin{align}\label{eq:Appparameter_identification}
   \frac{V'_\text{nn}}{\chi} = \frac{4\kappa^4}{\beta^4}\;, &&   \frac{\lambda'}{\chi} = \frac{2\kappa^2(\ell M_0)^2}{\beta^4}  \;.
\end{align}
To summarize, our implementation of the SG model is formally valid in the continuum limit (\mbox{$\ell M_0 =\kappa \sqrt{2\lambda'/V'_\text{nn}} \rightarrow 0$}), together with sufficiently large spin length ($J \rightarrow \infty$) and a large number of Rydberg atoms ($N\rightarrow \infty$). 
This requires tuning the atomic parameters according to Eqs.~\eqref{eq:Appparameter_identification}, which in principle allows to probe all regimes of the theory with \mbox{$0 < \beta^2 =2\kappa^2 \sqrt{\chi/V'_\text{nn}} < \infty$}. 

In practice, the range of parameters is mainly limited by the maximum value of $\chi \lesssim 2\pi\times 200$ kHz.
For example, $\kappa=1$, $J = 20$ and choosing $\chi = 2\pi\times 200$ kHz, $V'_\text{nn} = 2\pi\times 100$ kHz gives $\beta^2 \approx \pi$.
Moreover, for $\lambda' = 2\pi\times 5$ kHz, we then find $\ell M_0 \approx 0.3$. 
From these estimates, we thus expect that several tens of atoms are sufficient to simulate the sine-Gordon QFT with reasonable accuracy up to $\beta^2 \lesssim \pi$, that is deep in the massive phase.
Larger values of $\beta^2$ can be achieved by using the ponderomotive coupling with $\kappa>1$.
For example, taking $\kappa = 4$, $\chi = 2\pi\times 200$ kHz, $V'_\text{nn} = 2\pi\times 300$ kHz and $\lambda' = 2\pi\times 3$ kHz gives $\beta^2 \gtrsim 8\pi$ and $\ell M_0 \approx 0.6$.
With these parameters one can then investigate the critical properties of the model across the quantum phase transition.

\section{Free theory approximation \texorpdfstring{\\}{} for the SG model}
\label{Appendix:FreeTheory}

Let us consider the many-body Hamiltonian for the lattice SG theory:
\begin{align}
\hat{H}_{\text{SG}}^{(\text{lat})} = \sum_i & \left[\chi \hat{\pi}_i^2 -  \lambda' \cos(\hat{\varphi}_i) \right. \nonumber\\
& \left.  - \frac{V'_{\text{dd}}}{2}\sum_{j=1}^\infty \frac{1}{|j|^3}\cos \left(\hat{\varphi}_{i+j}-\hat\varphi_j\right) \right]\,.
\end{align}
In the small-fluctuations regime, which applies deep in the gapped phase of the model, we expand to quadratic order in $\hat \varphi_i$ and obtain the momentum space Hamiltonian
\begin{align}
\hat{H}_{\text{SG}}^{(\text{lat})} = & 
\sum_q \left( \chi \hat \pi_q \hat \pi_{-q} + \frac{1}{4\chi}\omega_q^2 \hat \varphi_q \hat \varphi_{-q} \right)\,,
\end{align}
where we defined
\begin{align}
\omega_q^2 &= 2\chi \left[\lambda^\prime + V^\prime_{\text{dd}}(\epsilon_0 - \epsilon_q) \right]\,,
\end{align}
and 
\begin{align}
\label{eq:long_range_ft}
\epsilon_q = \frac 1 2\sum_{j\neq 0} \frac{e^{\mr i qj}}{|j|^3}\,.
\end{align}
After performing the transformation \cite{Bruus2004}
\begin{align}
\label{eq:bogtrasform}
\hat \pi_q = \frac{i}{\sqrt{2}\ell_q} \left( \hat b^\dagger_{-q} - \hat b^{}_{q}\right)\,,\quad
\hat \varphi_q = \frac{\ell_q}{\sqrt{2}} \left( \hat b^\dagger_{-q} + \hat b^{}_{q}\right)\,,
\end{align}
with $\ell_q \equiv \sqrt{2\chi/\omega_q}$ and the bosonic operators satisfying $[\hat b^{}_{q},\hat b^\dagger_{q'}] = \delta_{q,q'}$ and $\hat b_{q}|\text{GS}\rangle = 0$, we obtain the diagonal Hamiltonian
\begin{align}
\hat{H}_{\text{SG}}^{(\text{lat})} = & \sum_q \omega_q \left(\hat b^\dagger_{q} \hat b^{}_{q} + \frac 1 2 \right)\,,
\end{align}
from which we deduce the excitation spectrum $\omega_q$ and the gap $\omega_0 = \sqrt{2\chi\lambda^\prime}$.

Having diagonalized the model within the quadratic approximation, we can now compute correlation functions of interest. For the sine-Gordon model a relevant correlator is, for example, represented by the expectation value of the vertex operator $\langle e^{\mr i\hat\varphi_i}\rangle$. By using Eq.~\eqref{eq:bogtrasform} and the Baker-Campbell-Hausdorff formula, we obtain
\begin{align}
\langle \cos\hat\varphi_i\rangle =\exp\left[ -\frac{\chi}{2N}\sum_q \frac{1}{\omega_q} \right] \underset{\lambda'\gg V_{\text{dd}}}{\longrightarrow} e^{-\frac{1}{2}\sqrt{\frac{\chi}{2\lambda'}}} \,.
\end{align}

Calculating the long-range effects of dipole-dipole interactions on the dispersion relation requires to evaluate the sum in Eq.~\eqref{eq:long_range_ft}, which is however slowly converging. An efficient method to compute $\epsilon_q$ exploits Ewald summation trick (see Ref.~\cite{Peter2015} and references therein for a description of the method). One therefore arrives at the expression
\begin{align}
\label{eq:ewald}
\epsilon_q =& \, t_0 \sum_G E_2\left( (q+G)^2 /4t_0 \right) + \nonumber \\
&\frac{t_0^{3/2}}{\sqrt{\pi}}  \sum_{j\neq0} E_{-1/2}(t_0 j^2) e^{\mr i q j} - \frac{2 t_0^{3/2}}{3\sqrt{\pi}}\,,
\end{align}
where $G=2\pi j'$, with $j'\in\mathbb Z$, are reciprocal lattice vectors, $t_0$ is a parameter that plays the role of a cut-off that we set to $t_0=1$ in the calculations, and we have also introduced the functions
\begin{align}
E_s(z) = \int_1^\infty \textrm d y \, y^{-s} e^{-z y}\,.
\end{align}
In the expression \eqref{eq:ewald}, only a few terms of each sum are needed as they are exponentially suppressed for $|j|, |j'| \gg 1$. The integrals are then efficiently computed by standard numerical methods. The final result for $\omega_q$ is shown in Fig.~\ref{fig:SG_illustration}b of the main text.

The small momenta behaviour of the dispersion can however be analytically obtained by considering the $G=0$ term and by using $E_2(z) = e^{-z} - z \Gamma(0,z)$, where $\Gamma(s,z) = \int_z^\infty \mathrm d t \, t^{s-1}e^{-t}$ is the incomplete Gamma function. The series expansion $\Gamma(0,z) = -\gamma - \log z + z +  O (z^2)$, with $\gamma$ the Euler-Mascheroni constant, therefore leads to 
\begin{align}
\epsilon_q \approx \epsilon_0 + c_2^{} q^2 + c_2^\prime\, q^2 \log q +  O (q^4)\,,
\end{align}
where $c_2 = \frac 1 4 (\gamma-1) - \frac{1}{\sqrt \pi} E_{-1/2}(1) -\log 2/2 \approx -0.739$ and $c_2^\prime = 1/2$. We have therefore found that the dispersion relation for $q\ll 1$ has a non-analytical correction scaling like $\sim q^2 \log q$. This is different from the 2D case \cite{Peter2015}, where a cusp-like non-analytical term $~\sim |q|$ instead provides the leading contribution and not the sub-leading one. 

\section{Implementing the massive SG model with large spins}
\label{Appendix:mSGimplementation}

In order to simulate the massive SG model in Eq.~\eqref{eq:H_QED}, we repeat a similar discretization as described for the SG model. 
However, this requires to include the MW field and the ponderomotive couplings simultaneously.
In particular, starting from the many-body Hamiltonian in Eq.~\eqref{eq:Hfull}, in the limit $J\rightarrow\infty$ we find
\begin{align}
\label{eq:AppQED_J}
    \hat{H}_{\text{mSG}}^{(\mr{lat})} = 
    \sum_i \Big[\chi \hat{\pi}_i^2 &-  \lambda'_\kappa \cos(\kappa\hat{\varphi}_i+\theta) -\Omega'\cos(\hat{\varphi}_i)\notag\\
    &-\sum_{j> i}\frac{V_{ij}'}{2}\cos \left(\hat{\varphi}_{j}-\hat\varphi_i\right) \Big] \;,
\end{align}
where we use the identification \mbox{$V_{ij}' = V_{ij} J(J+1)$}, $\Omega' = -\Omega_a [2\sqrt{J(J+1)}]$ and $\lambda'_\kappa = -2\lambda_\kappa [J(J+1)]^{\kappa/2}$, and neglect the Ising interactions. 
Our target model is again approached in the continuum limit with the identifications 
\begin{align}
    \frac{V'_\mr{nn}}{\chi} = \frac{\kappa^4}{(2\pi)^2} \,, &&
    \frac{\lambda'_\kappa}{\chi} =
    \frac{\kappa^2\ell^2\exp(\gamma)\Lambda m}{(2\pi)^2}\,, && 
    \frac{\Omega'}{\chi} = \frac{\kappa^4e^2 \ell^2}{(2\pi)^3}\,,
\end{align}
and keeping only nearest-neighbor terms $V'_\mr{nn}\equiv V'_{i,i+1}$. 

Let us briefly discuss the meaning of these identifications. 
To recover the model in Eq.~\eqref{eq:H_QED}, we need to approximate the interaction potential as 
$ - \Omega' \cos (\hat \varphi_i) - \lambda'_\kappa \cos(\kappa\hat \varphi_i) \rightarrow (\Omega'/2)\hat \varphi^2_i - \lambda'_\kappa \cos (\kappa\hat\varphi_i)$, up to a constant.
This requires $\Omega' \gg \lambda'_\kappa,\, \chi$ and $\kappa \gg 1$, which fixes the ratio $V'_\mr{nn}/\chi \gg 1$.
As $\Lambda \propto 1/\ell$ and $\Lambda \gg e$, we further obtain the requirement $V'_\mr{nn} \gg \Omega'$.
Together with the requirement $\Lambda \gg m$, these conditions can be satisfied simultaneously by tuning the couplings in the range 
\begin{align}\label{eq:Appsep_scales}
\chi,\, \lambda_\kappa' \ll \Omega' \ll V_\mr{nn}' \;,  \quad \lambda' \ll \sqrt{\chi V_\mr{nn}'}\,.
\end{align}

When these conditions are fulfilled, the lattice model $\hat{H}_{\text{mSG}}^{(\mr{lat})}$ faithfully reproduces $\hat{H}_{\text{mSG}}$ (Eq.~\eqref{eq:H_QED}), and thus the dual gauge theory with the coupling strength
\begin{align}
    \frac{e}{m} = \frac{\exp(\gamma) \pi}{\sqrt{2\pi}} \frac{\sqrt{\Omega'\chi}}{\lambda'_\kappa}\;,
\end{align}
where we used $\Lambda \ell = \pi$ \cite{jentsch2022physical}.
We emphasize that a suitable choice of parameters thus in principle allows us to probe the whole non-trivial range of the theory with $0 < e/m < \infty$.
For example, consider $\kappa = 5$, $J=30$, $\chi = 2\pi\times 50$ kHz, which gives $V'_\text{nn} \approx 2\pi\times 790$ kHz.
By further taking $\Omega' = 2\pi\times 200$ kHz and $\lambda' = 2\pi\times 50$ kHz, we obtain $e/m \approx 4.5$, while $e/\Lambda \approx 0.4$ and $m/\Lambda \approx 0.09$.

\section{State transfer protocol}
\label{Appendix:StateTransfer}

In the following Appendix we briefly summarize the underlying equations for the state transfer discussed in the main text Sec.~\ref{sec:ProspectQI}. 

For the sake of simplicity we consider here in the Appendix two harmonic oscillator modes $a$ and $b$ and we explicitly show how the quantum state from the oscillator $b$ is transferred to $a$. 
The Hamiltonian, generating the anticipated transfer dynamic is given by $\hat H = -V(\hat a^\dagger \hat b+ \hat b^\dagger \hat a)$. 
The Heisenberg equations for the creation operators read $\dot{\hat a}_h^\dagger = i[\hat H,\hat a_h^\dagger] = -iV \hat b_h^\dagger$ and $\dot{\hat b}_h^\dagger = i[\hat H,\hat b_h^\dagger] = -iV \hat a_h^\dagger$, where the subscript $h$ denotes the Heisenberg picture, and are solved by $\hat a_h^\dagger(t) = \cos(V t) \hat a^\dagger - i \sin(V t) \hat b^\dagger$ and $\hat b_h^\dagger(t) = \cos(V t) \hat b^\dagger - i \sin(V t) \hat a^\dagger$. Interestingly, the two oscillator modes swap at time $T = \pi/(2V)$, \emph{i.e.} $\hat a_h^\dagger(T) = -i\hat b^\dagger$ and $\hat b_h^\dagger(T) = -i\hat a^\dagger$, which will be important for state transfer as follows. 
To show the transfer we consider the initial state $\ket{\psi_0} = \sum_n d_n \ket{0,n}$ time evolved under the unitary operation $\hat U = \exp(-i\hat HT)$:
\begin{align}
    \hat U\ket{\psi_0} &=  \sum_n d_n \hat U(\hat b^\dagger)^n \hat U^\dagger \hat U \ket{0,0}/\sqrt{n!}\notag\\
    &= \sum_n d_n \hat b_h^\dagger(T) \ket{0,0}/\sqrt{n!} \\
    &= \sum_n d_n(-i)^n (\hat a^\dagger)^n \ket{0,0}/\sqrt{n!} = \sum_n(-i)^n  d_n \ket{n,0}\notag,
\end{align}
where the quantum state of the oscillator $b$ is completely transferred to $a$.
The generalization of this protocol to the Rydberg manifold is straightforward.

\bibliography{bib}

\end{document}